\documentclass[12pt,prd,aps,amssymb,amsmath,tightenlines,showpacs]{revtex4}
\usepackage{psfig}

\usepackage{graphics}
\usepackage[pdftex]{graphicx}
\newcommand{\half}{\mbox{$\textstyle\frac{1}{2}$}}
\begin{document}
\preprint{}

\title{Introduction to $\mathcal{PT}$-Symmetric Quantum Theory}

\author{Carl~M.~Bender\footnote{Permanent address: Department of Physics,
Washington University, St. Louis, MO 63130, USA.}}

\affiliation{Blackett Laboratory, Imperial College, London SW7 2BZ, UK}

\date{\today}

\begin{abstract}
In most introductory courses on quantum mechanics one is taught that the
Hamiltonian operator must be Hermitian in order that the energy levels be real
and that the theory be unitary (probability conserving). To express the
Hermiticity of a Hamiltonian, one writes $H=H^\dagger$, where the symbol
$\dagger$ denotes the usual Dirac Hermitian conjugation; that is, transpose and
complex conjugate. In the past few years it has been recognized that the
requirement of Hermiticity, which is often stated as an axiom of quantum
mechanics, may be replaced by the less mathematical and more physical
requirement of space-time reflection symmetry ($\mathcal{PT}$ symmetry) without
losing any of the essential physical features of quantum mechanics. Theories
defined by non-Hermitian $\mathcal{PT}$-symmetric Hamiltonians exhibit strange
and unexpected properties at the classical as well as at the quantum level.
This paper explains how the requirement of Hermiticity can be evaded and
discusses the properties of some non-Hermitian $\mathcal{PT}$-symmetric
quantum theories.
\end{abstract}


\maketitle

\section{Introduction}
\label{sec0}

The field of $\mathcal{PT}$-symmetric quantum theory is only six years old but
already hundreds of papers have been published on various aspects of $\mathcal{P
T}$-symmetric quantum mechanics and $\mathcal{PT}$-symmetric quantum field
theory. Three international conferences have been held (Prague, 2003; Prague,
2004; Shizuoka, 2004) and three more conferences are planned. Work on $\mathcal{
PT}$ symmetry began with the investigation of quantum-mechanical models and has
now extended into many areas including quasi-exact solvability, supersymmetry,
and quantum field theory. Recently, it has been recognized that there is a
connection between $\mathcal{PT}$-symmetric quantum mechanics and integrable
models. The aim of this paper is to introduce the subject at an elementary level
and to elucidate the properties of theories described by $\mathcal{P
T}$-symmetric Hamiltonians. This paper will make the field of $\mathcal{PT}$
symmetry accessible to students who are interested in exploring this exciting,
new, and active area of physics.

The central idea of $\mathcal{PT}$-symmetric quantum theory is to replace the
condition that the Hamiltonian of a quantum theory be Hermitian with the weaker
condition that it possess space-time reflection symmetry ($\mathcal{PT}$
symmetry). This allows us to construct and study many new kinds of Hamiltonians
that would previously have been ignored. These new Hamiltonians have remarkable
mathematical properties and it may well turn out that these new Hamiltonians
will be useful in describing the physical world. It is crucial, of course, that
in replacing the condition of Hermiticity by $\mathcal{PT}$ symmetry we do not
give up any of the key physical properties that a quantum theory must have. We
will see that if the $\mathcal{PT}$ symmetry of the Hamiltonian is not broken,
then the Hamiltonian will exhibit all of the features of a quantum theory
described by a Hermitian Hamiltonian. (The word {\it broken} as used here is a
technical term that will be explained in Sec.~\ref{sec1}.)

Let us begin by reviewing some basic ideas of quantum theory. For simplicity, in
this paper we restrict our attention to one-dimensional quantum-mechanical
systems. Also, we work in units where Planck's constant $\hbar=1$. In elementary
courses on quantum mechanics one learns that a quantum theory is specified by
the Hamiltonian operator that acts on a Hilbert space. The Hamiltonian $H$ does
three things:

\noindent\quad (i) The Hamiltonian determines the energy eigenstates $|E_n
\rangle$. These states are the eigenstates of the Hamiltonian operator and they
solve the time-independent Schr\"odinger equation $H|E_n\rangle=E_n|E_n\rangle$.
The energy eigenstates span the Hilbert space of physical state vectors. The
eigenvalues $E_n$ are the energy levels of the quantum theory. In principle, one
can observe or measure these energy levels. The outcome of such a physical
measurement is a real number, so it is essential that these energy eigenvalues
be real.

\noindent\quad (ii) The Hamiltonian $H$ determines the time evolution in the
theory. States $|t\rangle$ in the Schr\"odinger picture evolve in time according
to the time-dependent Schr\"odinger equation $H|t\rangle=-i\frac{d}{dt}|t\rangle
$, whose formal solution is $|t\rangle=e^{iHt}|0\rangle$. Operators $A(t)$ in
the Heisenberg picture evolve according to the time-dependent Schr\"odinger
equation $\frac{d}{dt}A(t)=-i[A(t),H]$, whose formal solution is $A(t)=e^{iHt}A(
0)e^{-iHt}$. 

\noindent\quad (iii) The Hamiltonian incorporates the symmetries of the theory.
A quantum theory may have two kinds of symmetries: continuous symmetries, such
as Lorentz invariance, and discrete symmetries, such as parity invariance and
time reversal invariance. A quantum theory is symmetric under a transformation
represented by an operator $A$ if $A$ commutes with the Hamiltonian that
describes the quantum theory: $[A,H]=0$. Note that if a symmetry transformation
is represented by a {\it linear} operator $A$ and if $A$ commutes with the
Hamiltonian, then the eigenstates of $H$ are also eigenstates of $A$. Two
important discrete symmetry operators are parity (space reflection), which is
represented by the symbol $\mathcal{P}$, and time reversal, which is represented
by the symbol $\mathcal{T}$. The operators $\mathcal{P}$ and $\mathcal{T}$ are
defined by their effects on the dynamical variables $\hat x$ (the position
operator) and $\hat p$ (the momentum operator). The operator $\mathcal{P}$ is
{\it linear} and has the effect of changing the sign of the momentum operator
$\hat p$ and the position operator $\hat x$: ${\hat p}\to-{\hat p}$ and ${\hat
x}\to-{\hat x}$. The operator $\mathcal{T}$ is antilinear and has the effect
${\hat p}\to-{\hat p}$, ${\hat x}\to{\hat x}$, and $i\to-i$. Note that $\mathcal
{T}$ changes the sign of $i$ because (like $\mathcal{P}$) $\mathcal T$ is
required to preserve the fundamental commutation relation $[{\hat x},{\hat p}]=
i$ of the dynamical variables in quantum mechanics.

Quantum mechanics is an association between states in a mathematical Hilbert
space and experimentally measurable probabilities. The norm of a vector in the
Hilbert space must be positive because this norm is a probability and a
probability must be real and positive. Furthermore, the inner product between
any two different vectors in the Hilbert space must be constant in time because
probability is conserved. The requirement that the probability not change with
time is called {\it unitarity}. Unitarity is a fundamental property of any
quantum theory and must not be violated.

To summarize the discussion so far, the two crucial properties of any quantum
theory are that the energy levels must be real and that the time evolution must
be unitary. There is a simple mathematical condition on the Hamiltonian that
guarantees the reality of the energy eigenvalues and the unitarity of the time
evolution; namely, that the Hamiltonian be {\it real and symmetric}. To explain
the term {\it symmetric}, as it is used here, let us first consider the
possibility that the quantum system has only a finite number of states. In this
case the Hamiltonian is a finite-dimensional symmetric matrix
\begin{equation}
H=\left(
\begin{array}{cccc}
a&b&c&\cdots\cr
b&d&e&\cdots\cr
c&e&f&\cdots\cr
\vdots&\vdots&\vdots&\ddots\cr
\end{array}
\right),
\label{eqmatrix}
\end{equation}
whose entries $a$, $b$, $c$, $d$, $e$, $f$, $\cdots$, are real numbers. For
systems having an infinite number of states we express $H$ in coordinate space
in terms of the dynamical variables $\hat x$ and $\hat p$. The $\hat x$ operator
in coordinate space is a {\it real} and {\it symmetric} diagonal matrix, all of
whose entries are the real number $x$. The $\hat p$ operator in coordinate space
is {\it imaginary} and {\it anti-symmetric} because ${\hat p}=-i\frac{d}{dx}$
when it acts to the right but, as we can see using integration by parts, ${\hat
p}$ changes sign ${\hat p}=i\frac{d}{dx}$ when it acts to the left. The operator
${\hat p}^2=-\frac{d^2}{dx^2}$ is real and symmetric. Thus, any Hamiltonian of
the form $H={\hat p}^2+V({\hat x})$ when written in coordinate space is real and
symmetric. (In this paper we use units in which $m=\half$ and we treat $\hat x$
and $\hat p$ as dimensionless.)

However, the condition that $H$ be real and symmetric is not the most general
condition that guarantees the reality of the energy levels and the unitarity of
the time evolution because it excludes the possibility that the Hamiltonian
matrix might be complex. Indeed, there are many physical applications which
require that the Hamiltonian be complex. There is a more general condition that
guarantees spectral reality and unitary time evolution and which includes real,
symmetric Hamiltonians as a special case. This condition is known as {\it
Hermiticity}. To express the condition that a complex Hamiltonian $H$ is
Hermitian we write $H=H^\dagger$. The symbol $\dagger$ represents {\it Dirac
Hermitian conjugation}; that is, combined transpose and complex conjugation. The
condition that $H$ must exhibit Dirac Hermiticity is often taught as an axiom of
quantum mechanics. The Hamiltonians $H={\hat p}^2+{\hat p}+V({\hat x})$ and $H={
\hat p}^2+{\hat p}{\hat x}+{\hat x}{\hat p}+V({\hat x})$ are complex and
nonsymmetric but they are Hermitian.

In this paper we show that while Hermiticity is sufficient to guarantee the two
essential properties of quantum mechanics, it is not necessary. We describe here
an alternative way to construct complex Hamiltonians that still guarantees the
reality of the eigenvalues and the unitarity of time evolution and which also
includes real, symmetric Hamiltonians as a special case. We will maintain the
symmetry of the Hamiltonians in coordinate space, but we will allow the matrix
elements to become complex in such a way that the condition of space-time
reflection symmetry ($\mathcal{PT}$ symmetry) is preserved. The new kinds of
Hamiltonians discussed in this paper are symmetric and have the property that 
they commute with the $\mathcal{PT}$ operator: $[H,\mathcal{PT}]=0$. In analogy
with the property of Hermiticity $H=H^\dagger$, we will express the property
that a Hamiltonian is $\mathcal{PT}$ symmetric by using the notation $H=H^
\mathcal{PT}$. We emphasize that our new kinds of complex Hamiltonians are
symmetric in coordinate space but are not Hermitian in the Dirac sense. To
reiterate, acceptable complex Hamiltonians may be either Hermitian $H=H^\dagger$
or $\mathcal{PT}$-symmetric $H=H^\mathcal{PT}$, but not both. Real symmetric
Hamiltonians may be both Hermitian and $\mathcal{PT}$-symmetric.
 
Using $\mathcal{PT}$ symmetry as an alternative condition to Hermiticity, we can
construct infinitely many new Hamiltonians that would have been rejected in the
past because they are not Hermitian. An example of such a $\mathcal{PT
}$-symmetric Hamiltonian is
\begin{equation}
H={\hat p}^2+i{\hat x}^3.
\label{eq1}
\end{equation}
We do not regard the condition of Hermiticity as wrong. Rather, the condition of
$\mathcal{PT}$ symmetry offers the possibility of studying new quantum theories
that may even describe measurable physical phenomena. Indeed, non-Hermitian
$\mathcal{PT}$-symmetric Hamiltonians have already been used to describe such
phenomena as the ground state of a quantum system of hard spheres \cite{rf30},
Reggeon field theory \cite{rf2}, and the Lee-Yang edge singularity \cite{rf1}.
Although at the time that they were written these papers were criticized for
using Hamiltonians that were not Hermitian, we now understand that these
Hamiltonians have spectral positivity and that the associated quantum theories
are unitary because these Hamiltonians are $\mathcal{PT}$-symmetric. In physics
we should keep an open mind regarding the kinds of theories that we are willing
to consider. Gell-Mann's ``totalitarian principle'' states that among the
possible physical theories ``Everything which is not forbidden is compulsory.''

This paper is organized as follows: I discuss in a personal way my discovery of
$\mathcal{PT}$-symmetric quantum mechanics and give a brief history of the early
days of this subject in Sec.~\ref{sec1}. Section~\ref{sec2} explains how to
calculate the energy levels of a $\mathcal{PT}$-symmetric Hamiltonian. Section
\ref{sec3} describes the classical mechanics of $\mathcal{PT}$-symmetric
Hamiltonians. Next, in Sec.~\ref{sec4} we show that a Hamiltonian having an
unbroken $\mathcal{PT}$ symmetry defines a {\it unitary} quantum theory. The
demonstration of unitarity is based on showing that $\mathcal{PT}$-symmetric
Hamiltonians that have an unbroken $\mathcal{PT}$ symmetry also possess a new
parity-like symmetry; this symmetry is represented by a new operator that we
call $\mathcal{C}$. We give in Sec.~\ref{sec5} a simple $2\times2$ matrix
illustration of the procedures used in Sec.~\ref{sec4}. In Sec.~\ref{sec6} we
discuss the nature of observables in $\mathcal{PT}$-symmetric quantum-mechanical
theories. We show how to calculate the $\mathcal{C}$ operator in
Sec.~\ref{sec7}. In Sec.~\ref{sec8} we explain why one may regard $\mathcal{P
T}$-symmetric quantum mechanics as a complex version of ordinary quantum
mechanics. Finally, in Sec.~\ref{sec9} we discuss some possible physical
applications of $\mathcal{PT}$-symmetric quantum mechanics.

\section{A Personal History of $\mathcal{PT}$ Symmetry}
\label{sec1}

My first encounter with a non-Hermitian complex Hamiltonian dates back to the
summer of 1993. In the course of a private conversation with D. Bessis at CEN
Saclay, I learned that he and J. Zinn-Justin had noticed that the eigenvalues of
the Hamiltonian operator in (\ref{eq1}) seemed to be real and they wondered if
the spectrum (the set of energy eigenvalues of the Hamiltonian) might be
entirely real. (Their interest in the Hamiltonian (\ref{eq1}) was inspired by
early work on the Lee-Yang edge singularity \cite{rf1}.) At the time I did not
plan to pursue this conjecture further because it seemed absurd that a complex
non-Hermitian Hamiltonian might have real energy levels.

I did not know it at the time, but Bessis and Zinn-Justin were not the first to
notice that a complex cubic quantum-mechanical Hamiltonian might have real
eigenvalues. For example, early studies of Reggeon field theory in the late
1970's led a number of investigators to observe that model cubic
quantum-mechanical Hamiltonians like that in (\ref{eq1}) might have real
eigenvalues \cite{rf2}. Also, E.~Caliceti {\it et al.}~observed in 1980 that on
the basis of Borel summability arguments the spectrum of a Hamiltonian related
to (\ref{eq1}) is real \cite{rf7}. In each of these cases the possibility that a
complex non-Hermitian Hamiltonian might have real energy levels was viewed as an
isolated curiosity. It was believed that such a Hamiltonian could not describe a
valid theory of quantum mechanics because the non-Hermiticity of the Hamiltonian
would result in nonunitary time evolution \cite{rf1}.

I did not forget about the remarkable Hamiltonian in (\ref{eq1}) and in 1997 I
decided to investigate it. I suspected that if the spectrum of this Hamiltonian
was real, it was probably due to the presence of a symmetry and I realized that
(\ref{eq1}) does possess $\mathcal{PT}$ symmetry because any real function of $i
{\hat x}$ is $\mathcal{PT}$-symmetric. I decided that a simple and natural way
to determine the spectrum of (\ref{eq1}) would be to use the delta expansion, a
perturbative technique that I had developed several years earlier for solving
nonlinear problems \cite{rf3}. I asked my former graduate student S.~Boettcher
to join me in this investigation.

The delta expansion is an extremely simple technique for solving nonlinear
problems perturbatively  (approximately). The idea of the delta expansion is to
introduce a small perturbation parameter $\delta$ into a nonlinear problem in
such a way that $\delta$ is a measure of the nonlinearity of the problem. To
illustrate how the delta expansion is used, consider the Thomas-Fermi equation,
a difficult nonlinear boundary-value problem that describes the approximate
electric charge distribution in an atom:
\begin{equation}
y''(x)=y^{3/2}x^{-1/2},\quad y(0)=1, \quad y(\infty)=0.
\label{eq2}
\end{equation}
There is no exact closed-form solution to this problem. However, a nice way to
solve this problem perturbatively is to introduce the parameter $\delta$ in the
exponent:
\begin{equation}
y''(x)=y(y/x)^\delta,\quad y(0)=1, \quad y(\infty)=0.
\label{eq3}
\end{equation}
When $\delta=\half$, (\ref{eq3}) reduces to (\ref{eq2}). However, in (\ref{eq3})
we treat the parameter $\delta$ as small $(\delta\ll 1)$. When $\delta=0$, the
problem becomes {\it linear} and therefore it can be solved exactly; the
solution is $y_0(x)=e^{-x}$. This is the leading term in the perturbation
expansion for $y(x)$, which has the form $y(x)=\sum_{n=0}^\infty y_n(x)\delta^n
$. It is easy to calculate the coefficients of the higher powers of $\delta$. At
the end of the calculation one sets $\delta=1/2$, and from just the first few
terms in the perturbation series one obtains a good numerical approximation to
the solution to the Thomas-Fermi equation.

I was eager to find out what would happen if we applied delta-expansion methods
to (\ref{eq1}). We replaced the Hamiltonian (\ref{eq1}) by the one-parameter
family of Hamiltonians
\begin{equation}
H={\hat p}^2+{\hat x}^2(i{\hat x})^\delta,
\label{eq4}
\end{equation}
where $\delta$ is regarded as a small real parameter. There are two advantages 
in inserting $\delta$ in this fashion: First, the new Hamiltonian remains
$\mathcal{PT}$-symmetric for {\it all} real $\delta$. Thus, the insertion of
$\delta$ maintains the $\mathcal{PT}$ symmetry of the original problem. Second,
when $\delta=0$, the Hamiltonian (\ref{eq4}) reduces to that of the harmonic
oscillator, which can be solved exactly because the underlying classical
equations of motion are linear. Each of the energy levels of (\ref{eq4})
have a delta expansion of the general form
\begin{equation}
E=\textstyle{\sum_{n=0}^\infty}a_n\delta^n.
\label{eq5}
\end{equation}
The series coefficients are easy to calculate and they are real. Thus, assuming
that this delta expansion converges, the eigenvalues of $H$ in (\ref{eq4})
must be real. At the end of the calculation we set $\delta=1$ in (\ref{eq5}) to
recover the eigenvalues of the original Hamiltonian in (\ref{eq1}).

The problem with the delta expansion is that it is difficult to prove rigorously
that the expansion converges. We were able to conclude only that for every $n$
there is always a neighborhood about $\delta=0$ in which the delta expansion for
the first $n$ eigenvalues of (\ref{eq4}) converges and thus these eigenvalues
are real when $\delta$ is real. Our discovery that the first $n$ eigenvalues of
the complex Hamiltonian (\ref{eq4}) were real for a small range of real $\delta$
near $\delta=0$ was astonishing to us, but we were disappointed that the delta
expansion was not powerful enough to determine whether all of the eigenvalues of
the original complex Hamiltonian (\ref{eq1}) are real. However, our
delta-expansion analysis inspired us to perform detailed perturbative and
numerical studies of the spectrum of $H$ in (\ref{eq4}). To our amazement we
found that {\it all of the eigenvalues of $H$ remain real for all} $\delta\geq0$
\cite{rf4}. We coined the term $\mathcal{PT}$-symmetric to describe these new
non-Hermitian complex Hamiltonians having real energy levels \cite{rf4.5}.

To present the results of our numerical studies, we rewrite the Hamiltonian
(\ref{eq4}) as
\begin{equation}
H={\hat p}^2-(i{\hat x})^N,
\label{eq6}
\end{equation}
where $N$ is a continuous real parameter \cite{rfen}. The eigenvalues of this
Hamiltonian are entirely real for all $N\geq2$, while for $N<2$ the spectrum is
partly real and partly complex. Clearly, the Hamiltonian $H={\hat p}^2+i{\hat x}
^3$ in (\ref{eq1}) is just one member of a huge and remarkable class of complex
Hamiltonians whose energy levels are real and positive. The spectrum of $H$
exhibits three distinct behaviors as a function of $N$ (see Fig.~\ref{figcmb1}):

\begin{figure}[b!]\vspace{2.3in}
\includegraphics{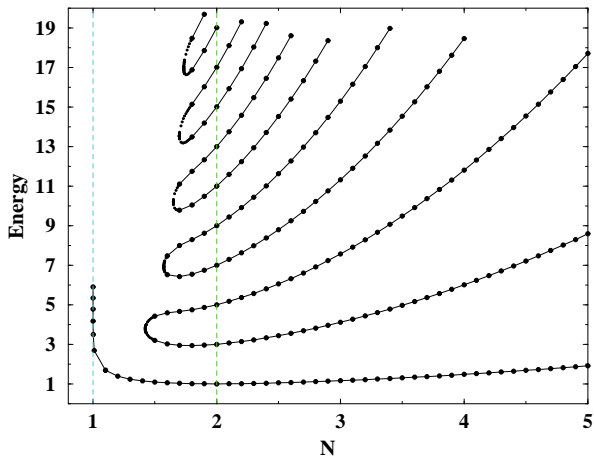}
\caption{Real energy levels of the Hamiltonian $H={\hat p}^2-(i{\hat x})^N$ as a
function of the parameter $N$. When $N\geq2$, the entire spectrum is real and
positive. The lower bound of this region, $N=2$, corresponds to the harmonic
oscillator, whose energy levels are $E_n=2n+1$. When $1<N<2$, there are a finite
number of positive real eigenvalues and infinitely many complex conjugate pairs
of eigenvalues. As $N$ decreases from $2$ to $1$, the number of real eigenvalues
decreases and when $N\leq1.42207$, the only real eigenvalue is the ground-state
energy. As $N\to1^+$, the ground-state energy diverges. For $N\leq1$ there are
no real eigenvalues.}
\label{figcmb1}
\end{figure}

(i) When $N\geq2$ the spectrum is infinite, discrete, and entirely real and
positive. This region includes the case $N=4$ for which $H={\hat p}^2-{\hat
x}^4$. Amazingly, the spectrum of this wrong-sign potential is positive and
discrete. (Also, $\langle{\hat x}\rangle\neq 0$ in the ground state because $H$ 
breaks parity symmetry!) At the lower bound $N=2$ of this region lies the
harmonic oscillator. (ii) A transition occurs at $N=2$. When $1<N<2$ there are
only a {\it finite} number of positive real eigenvalues and an infinite number
of complex conjugate pairs of eigenvalues. We say that in this region the
$\mathcal{PT}$ symmetry is {\it broken} and that $N\geq2$ is a region of {\it
unbroken} $\mathcal{PT}$ symmetry. (We explain the notion of broken and unbroken
$\mathcal{PT}$ symmetry in greater detail below.) As $N$ decreases from $2$ to
$1$, adjacent energy levels merge into complex conjugate pairs beginning at the
high end of the spectrum. Ultimately, the only remaining real eigenvalue is the
ground-state energy, which diverges as $N\to1^+$ \cite{rf5}. (iii) When $N\leq1$
there are no real eigenvalues.

It is apparent that the reality of the spectrum of (\ref{eq6}) when $N\geq2$ is
connected with its $\mathcal{PT}$ symmetry. The association between $\mathcal{P
T}$ symmetry and the reality of spectra can be understood as follows: We say
that the $\mathcal{PT}$ symmetry of a Hamiltonian $H$ is {\it unbroken} if all
of the eigenfunctions of $H$ are simultaneously eigenfunctions of $\mathcal{P
T}$ \cite{foot1}.

Here is a proof that if the $\mathcal{PT}$ symmetry of a Hamiltonian $H$ is
unbroken, then the spectrum of $H$ is real: Assume that a Hamiltonian $H$
possesses $\mathcal{PT}$ symmetry (i.e., that $H$ commutes with the $\mathcal{P
T}$ operator) and that if $\phi$ is an eigenstate of $H$ with eigenvalue $E$,
then it is simultaneously an eigenstate of $\mathcal{PT}$ with eigenvalue
$\lambda$:
\begin{equation}
H\phi=E\phi\quad{\rm and}\quad\mathcal{PT}\phi=\lambda\phi.
\label{eq7}
\end{equation}

We begin by showing that the eigenvalue $\lambda$ is a pure phase. Multiply
$\mathcal{PT}\phi=\lambda\phi$ on the left by $\mathcal{PT}$ and use the fact
that $\mathcal{P}$ and $\mathcal{T}$ commute and that $\mathcal{P}^2=\mathcal{T}
^2=1$ to conclude that $\phi=\lambda^*\lambda\phi$ and thus $\lambda=e^{i\alpha}
$ for some real $\alpha$. Next, introduce a convention that we use throughout
this paper. Without loss of generality we replace the eigenfunction $\phi$ by
$e^{-i\alpha/2}\phi$ so that its eigenvalue under the operator $\mathcal{PT}$ is
unity:
\begin{equation}
\mathcal{PT}\phi=\phi.
\label{eq8}
\end{equation}
Next, we multiply the eigenvalue equation $H\phi=E\phi$ on the left by
$\mathcal{PT}$ and use $[\mathcal{PT},H]=0$ to obtain $E\phi=E^*\phi$. Hence,
$E=E^*$ and the eigenvalue $E$ is real.

The crucial assumption in this argument is that $\phi$ is simultaneously an
eigenstate of $H$ and of $\mathcal{PT}$. In quantum mechanics if a linear
operator $X$ commutes with the Hamiltonian $H$, then the eigenstates of $H$ are
also eigenstates of $X$. However, we emphasize that the operator $\mathcal{PT}$
is {\it not linear} (it is antilinear) and thus we must make the extra
assumption that the $\mathcal{PT}$ symmetry of $H$ is unbroken; that is, $\phi$
is simultaneously an eigenstate of $H$ and $\mathcal{PT}$. This extra assumption
is nontrivial because it is hard to determine {\it a priori} whether the
$\mathcal{PT}$ symmetry of a particular Hamiltonian $H$ is broken or unbroken.
For $H$ in (\ref{eq6}) the $\mathcal{PT}$ symmetry is unbroken when $N\geq2$ and
it is broken when $N<2$. The conventional Hermitian Hamiltonian for the
quantum-mechanical harmonic oscillator lies at the boundary of the unbroken and
the broken regimes.

I am delighted at the research activity that my work has inspired. In 2001 Dorey
{\it et al.}~proved rigorously that the spectrum of $H$ in (\ref{eq6}) is real
and positive \cite{rf6} in the region $N\geq2$. Dorey {\it et al.}~used
techniques such as the Bethe {\it ansatz} and the Baxter-TQ relation, which are
used in the study of integrable models and conformal quantum field theory. In
doing so they have helped to establish a remarkable connection between the
ordinary differential equation (the Schr\"odinger equation) that describes
$\mathcal{PT}$-symmetric quantum mechanics and the study of integrable models.
This connection, which has become known as the ODE/IM correspondence, is rich
and profound and will lead to a much deeper understanding of both types of
theories. Many other $\mathcal{PT}$-symmetric Hamiltonians for which space-time
reflection symmetry is not broken have been investigated, and the spectra of
these Hamiltonians have also been shown to be real and positive
\cite{rf7,rf8,rf9,rf10}. Evidently, the phenomenon of $\mathcal{PT}$ symmetry is
quite widespread and arises in many contexts.

\section{Energy Levels of a $\mathcal{PT}$-Symmetric Hamiltonian}
\label{sec2}

The purpose of this section is to explain how to calculate the eigenvalues of
the complex Hamiltonian operator in (\ref{eq5}). To calculate the energy levels
of a $\mathcal{PT}$-symmetric Hamiltonian we adopt the techniques that are used
for calculating the energy levels of conventional Hermitian Hamiltonians. These
techniques involve converting the formal eigenvalue problem $H\phi=E\phi$ to a
Schr\"odinger differential equation whose solutions satisfy appropriate boundary
conditions. This Schr\"odinger equation is then solved numerically or by using
approximate methods such as WKB.

The Schr\"odinger eigenvalue problem for the $\mathcal{PT}$-symmetric
Hamiltonian (\ref{eq6}) is
\begin{eqnarray}
-\phi_n''(x)-(ix)^N\phi_n(x)=E_n\phi_n(x),
\label{eq9}
\end{eqnarray}
where $E_n$ is the $n$th eigenvalue. For a Hermitian Hamiltonian the boundary
conditions that give quantized energy levels $E_n$ are that the eigenfunctions
$\phi_n(x)\to0$ as $|x|\to\infty$ on the real axis. This condition suffices for
(\ref{eq9}) when $1<N<4$, but for $N\geq4$ we must continue the eigenvalue
problem for (\ref{eq9}) into the complex-$x$ plane. Thus, we replace the
real-$x$ axis by a contour in the complex plane along which the differential
equation holds. The boundary conditions that lead to quantization are imposed at
the endpoints of this contour. (The rest of this brief section is somewhat
technical and may be skipped by those without a background in complex
differential equations. See Ref.~\cite{rf11} for more information on how to
solve such problems.)

\begin{figure}[t!]\vspace{2.0in}
\includegraphics{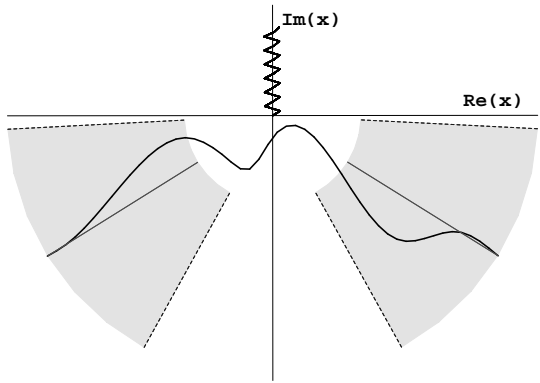}
\caption{Wedges in the complex-$x$ plane containing the contour on which the
eigenvalue problem for the differential equation (\ref{eq9}) for $N=4.2$ is
posed. In these wedges $\phi(x)$ vanishes exponentially as $|x|\to\infty$. The
wedges are bounded by lines along which the solution to the differential
equation is oscillatory.}
\label{figcmb2}
\end{figure}

The endpoints of this contour lie in regions in the complex-$x$ plane in which 
the eigenfunctions $\phi_n(x)\to0$ exponentially as $|x|\to\infty$. These
regions are known as {\it Stokes wedges} (see Fig.~\ref{figcmb2}). The Stokes
wedges are bounded by the lines along which the solution to the differential
equation is oscillatory \cite{rf12}. There are many wedges in which we can
require that $\phi(x)\to0$ as $|x|\to\infty$. Thus, there are many eigenvalue
problems associated with a given differential equation \cite{rf11}. For a given
value of $N$ we must first identify which one of these eigenvalue problems is
associated with (\ref{eq9}). To do so we start with the harmonic oscillator
problem at $N=2$ and smoothly vary the parameter $N$ until it reaches the
given value. At $N=2$ the eigenfunctions vanish in wedges of angular opening
$\half\pi$ centered about the negative-real and positive-real $x$ axes. For any
$N\geq1$ the centers of the left and right wedges lie at the angles
\begin{eqnarray}
\theta_{\rm left}=-\pi+\textstyle{\frac{N-2}{2N+4}}\pi\quad{\rm and}\quad
\theta_{\rm right}=-\textstyle{\frac{N-2}{2N+4}}\pi.
\label{eq10}
\end{eqnarray}
The opening angle of these wedges is $\Delta=\frac{2}{N+2}\pi$. The differential
equation (\ref{eq9}) may be integrated on any path in the complex-$x$ plane so
long as the path approaches complex infinity inside the left wedge and inside
the right wedge. These wedges contain the real-$x$ axis when $1<N<4$.

As $N$ increases from $2$, the left and right wedges rotate downward into the
complex-$x$ plane and become thinner. We can see on Fig.~\ref{figcmb1} that the
eigenvalues grow with $N$ as $N\to\infty$. At $N=\infty$ the differential equation contour runs up and down the negative imaginary axis, and this leads to
an interesting limiting eigenvalue problem. Because all of the eigenvalues of
(\ref{eq6}) diverge like $N^2$ as $N\to\infty$, for large $N$ we replace $H$ by
the rescaled Hamiltonian $H/N^2$. In the limit $N\to\infty$ this new Hamiltonian
becomes exactly solvable in terms of Bessel functions. The eigenvalue problem at
$N=\infty$ is the $\mathcal{PT}$-symmetric equivalent of the square well in
ordinary Hermitian quantum mechanics \cite{rf12.5}.

As $N$ decreases below $2$, the wedges become wider and rotate into the
upper-half $x$ plane. At $N=1$ the angular opening of the wedges
is ${2\over3}\pi$ and the wedges are centered at ${5\over6}\pi$ and ${1\over6}
\pi$. Thus, the wedges become contiguous at the positive-imaginary $x$ axis, and
the differential equation contour can be pushed off to infinity. Hence, there is
no eigenvalue problem when $N=1$ and, as we would expect, the ground-state
energy diverges as $N\to1^+$ (see Fig.~\ref{figcmb1}).

Having defined the eigenvalue problem for the Hamiltonian in (\ref{eq6}), we can
solve the differential equation by using numerical methods. We can also can use
approximate analytical methods such as WKB \cite{rf12}. WKB gives a good
approximation to the eigenvalues in Fig.~\ref{figcmb1} when $N\geq2$. The
novelty of this WKB calculation is that it must be performed in the complex
plane. The turning points $x_{\pm}$ are those roots of $E+(ix)^N=0$ that {\sl
analytically continue} off the real axis as $N$ moves away from $N=2$:
\begin{eqnarray}
x_-=E^{1/N}e^{i\pi(3/2-1/N)},\quad x_+=E^{1/N}e^{-i\pi(1/2-1/N)}.
\label{eq11}
\end{eqnarray}
These points lie in the lower (upper) $x$ plane in Fig.~\ref{figcmb2} when $N>2$
($N<2$).

The WKB quantization condition is $(n+1/2)\pi=\int_{x_-}^{x_+}dx\,\sqrt{E+(ix)^N
}$. It is crucial that the integration path be such that this {\it integral is
real.} When $N>2$ this path lies entirely in the lower-half $x$ plane, and when
$N=2$ the path lies on the real axis. When $N<2$ the path is in the upper-half
$x$ plane; it crosses the cut on the positive-imaginary axis and thus is {\it
not a continuous path joining the turning points.} Hence, WKB fails when $N<2$.

When $N\geq2$, the WKB calculation gives
\begin{eqnarray}
E_n\sim\left[{\Gamma(3/2+1/N)\sqrt{\pi}(n+1/2)\over\sin(\pi/N)\Gamma(1+1/N)}
\right]^{2N/(N+2)}\quad(n\to\infty).
\label{eq12}
\end{eqnarray}
This result is quite accurate. The fourth exact eigenvalue (obtained using
Runge-Kutta) for the case $N=3$ is 11.3143 while WKB gives 11.3042, and the
fourth exact eigenvalue for the case $N=4$ is 18.4590 while WKB gives 18.4321.

\section{$\mathcal{PT}$-Symmetric Classical Mechanics}
\label{sec3}

In the study of classical mechanics the objective is to describe the motion
of a particle satisfying Newton's law $F=ma$. The trajectory $x(t)$ of the
particle is a {\it real} function of time $t$. The classical equation of motion
for the complex $\mathcal{PT}$-symmetric Hamiltonian (\ref{eq6}) describes a
particle of energy $E$ subject to {\it complex} forces. Thus, we have the
surprising result that classical $\mathcal{PT}$-symmetric Hamiltonians describe
motion that is not limited to the real-$x$ axis. The classical path $x(t)$ may
lie in the complex-$x$ plane. The purpose of this section is to describe this
remarkable possibility \cite{BBM}.

An intriguing aspect of Fig.~\ref{figcmb1} is the transition at $N=2$. As $N$
goes below $2$, the eigenvalues begin to merge into complex conjugate pairs. The
onset of eigenvalue merging can be thought of as a phase transition. We show in
this section that the underlying cause of this quantum transition can be
understood by studying the theory at {\sl classical} level.

\begin{figure}[t!]
\vspace{1.6in}
\includegraphics{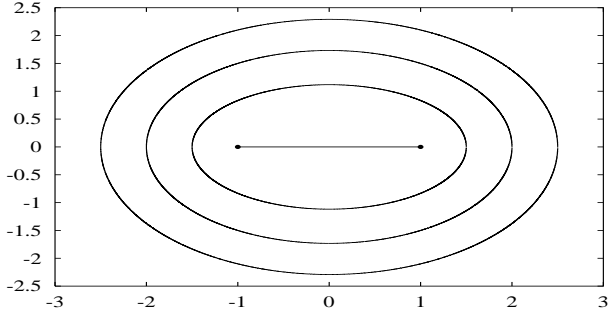}
\caption{Classical paths in the complex-$x$ plane for the $N=2$ oscillator. The
paths form a set of nested ellipses. These closed periodic orbits occur when
the $\mathcal{PT}$ symmetry is unbroken.}
\label{figpeter1}
\end{figure}

The trajectory $x(t)$ of a classical particle governed by the $\mathcal{PT
}$-symmetric Hamiltonian (\ref{eq6}) obeys $\pm dx[E+(ix)^N]^{-1/2}=2dt$. While
$E$ and $dt$ are real, $x(t)$ lies in the complex plane in Fig.~\ref{figcmb2}.
When $N=2$ (the harmonic oscillator), there is one classical path that
terminates at the classical turning points $x_\pm$ in (\ref{eq11}). Other paths
are nested ellipses with foci at the turning points (see Fig.~\ref{figpeter1}).
All these paths have the same period.

\begin{figure}[t!]
\vspace{1.6in}
\includegraphics{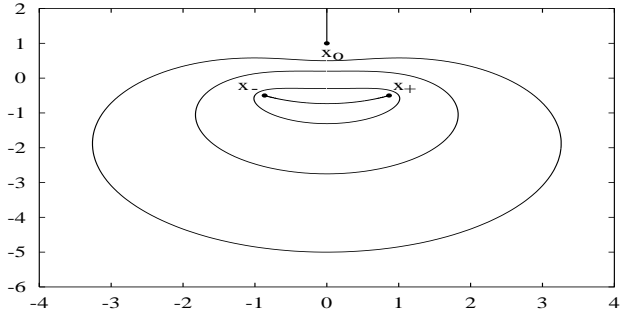}
\caption{Classical paths in the complex-$x$ plane for the $N=3$ oscillator. In
addition to the periodic orbits, one path runs off to $i\infty$ from the turning
point on the imaginary axis.}
\label{figpeter2}
\end{figure}

\begin{figure}[t!]
\vspace{1.6in}
\includegraphics{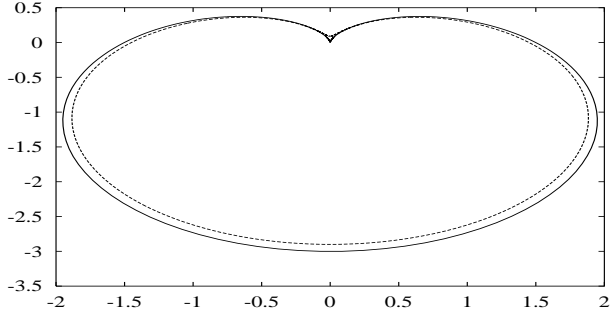}
\caption{Classical paths in the complex-$x$ plane for the $N=3$ oscillator. As
the paths get larger, they approach a shape resembling a cardioid. We have
plotted the {\it rescaled} paths.}
\label{figpeter3}
\end{figure}

\begin{figure}[t!]
\includegraphics{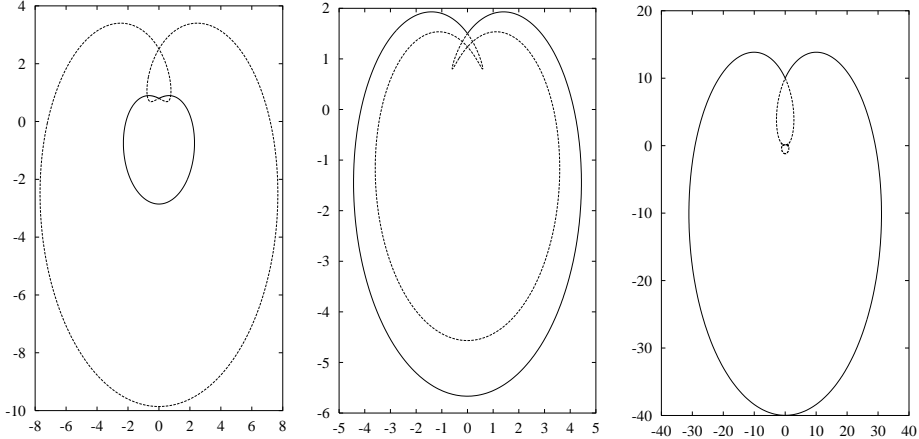}
\caption{Classical paths for the case $N=2.5$. These paths do not intersect. The
graph shows the projection of the parts of the path that lie on three different
sheets of the Riemann surface. As the size of the paths increases, a limiting
cardioid appears on the principal sheet. On the remaining sheets of the surface
the path exhibits a knot-like topological structure.}
\label{figpeter7c}
\end{figure}

When $N=3$, there is again a classical path that joins the left and right
turning points and an infinite class of paths enclosing the turning points (see
Fig.~\ref{figpeter2}). As these paths increase in size, they approach a cardioid
shape (see Fig.~\ref{figpeter3}). The indentation in the limiting cardioid
occurs because paths may not cross, and thus all periodic paths must avoid the
path in Fig.~\ref{figpeter2} that runs up the imaginary axis. When $N$ is
noninteger, we obtain classical paths that move off onto {\it different sheets}
of the Riemann surface (see Fig.~\ref{figpeter7c}).

In general, whenever $N\geq2$, the trajectory joining $x_\pm$ is a smile-shaped
arc in the lower complex plane. The motion is {\sl periodic}. Thus, the equation
describes a {\it complex pendulum} whose (real) period $T$ is given by
\begin{eqnarray}
T=2E^{2-N\over2N}\cos\left[{(N-2)\pi\over2N}\right]{\Gamma(1+1/N)\sqrt{\pi}\over
\Gamma(1/2+1/N)}.
\label{eq13}
\end{eqnarray}

\begin{figure}[t!]
\includegraphics{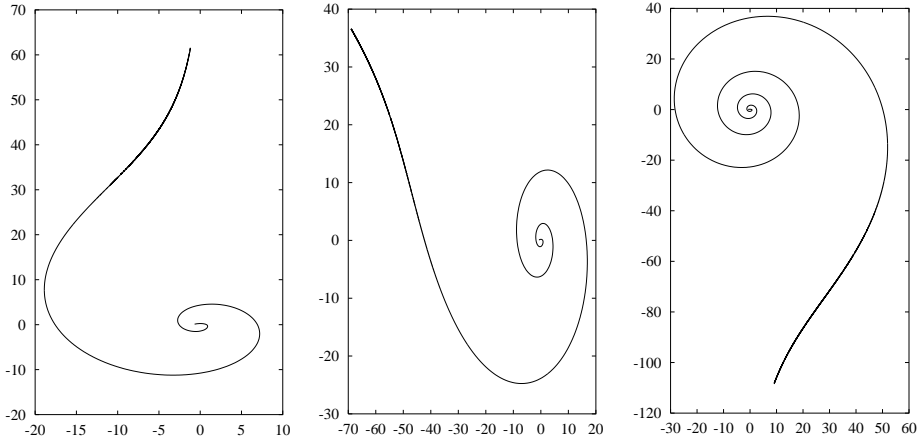}
\caption{Classical paths in the complex-$x$ plane for $N=1.8$, $N=1.85$ and
$N=1.9$. These nonperiodic paths spiral outward to infinity. As $N\to2$ from
below, the number of turns in the spiral increases. The lack of periodic orbits
corresponds to a broken $\mathcal{PT}$ symmetry.}
\label{figpeter9c}
\end{figure}

Below the transition at $N=2$ a path starting at one turning point, say $x_-$,
moves toward but {\sl misses} the turning point $x_+$. This path spirals
outward, crossing from sheet to sheet on the Riemann surface, and eventually
veers off to infinity. Hence, the period abruptly becomes infinite. The total
angular rotation of the spiral is finite for all $N<2$, but it becomes infinite
as $N\to2^-$ (see Fig.~\ref{figpeter9c}).

\section{$\mathcal{PT}$-Symmetric Quantum Mechanics}
\label{sec4}

The discovery that the eigenvalues of many $\mathcal{PT}$-symmetric Hamiltonians
are real and positive raises an urgent question: Does a non-Hermitian
Hamiltonian such as $H$ in (\ref{eq6}) define a physical theory of quantum
mechanics or is the positivity of the spectrum merely an intriguing mathematical
property of special classes of complex eigenvalue problems? A {\it physical}
quantum theory (i) must possess a Hilbert space of state vectors and this
Hilbert space must have an inner product with a positive norm; (ii) the time
evolution of the theory must be unitary; that is, the norm must be preserved in
time.

A definitive answer to this question has been found \cite{rf14,rf13}. For a
complex non-Hermitian Hamiltonian having an {\it unbroken} $\mathcal{PT}$
symmetry, a linear operator $\mathcal{C}$ that commutes with both $H$ and
$\mathcal{PT}$ can be constructed. We denote the operator representing this
symmetry by $\mathcal{C}$ because the properties of $\mathcal{C}$ are similar to
those of the charge conjugation operator in ordinary particle physics. The inner
product with respect to $\mathcal{CPT}$ conjugation is
\begin{equation}
\langle\psi|\chi\rangle^{\mathcal{CPT}}=\textstyle{\int}dx\,\psi^{
\mathcal{CPT}}(x)\chi(x),
\label{eq14}
\end{equation}
where $\psi^{\mathcal{CPT}}(x)=\int dy\,{\mathcal{C}}(x,y)\psi^*(-y)$. This
inner product satisfies the requirements for the quantum theory defined by $H$
to have a Hilbert space with a positive norm and to be a unitary theory of
quantum mechanics.

To explain the construction of the $\mathcal{C}$ operator we begin by
summarizing the mathematical properties of the solution to the eigenvalue
problem (\ref{eq9}) associated with the Hamiltonian $H$ in (\ref{eq6}). Recall
from Fig.~\ref{figcmb2} that this differential equation is imposed on an
infinite contour in the complex-$x$ plane and that for large $|x|$ this contour
lies in wedges placed symmetrically with respect to the imaginary-$x$ axis as in
Fig.~\ref{figcmb2}. When $N\geq2$, $H$ has an unbroken $\mathcal{PT}$ symmetry.
Thus, the eigenfunctions $\phi_n(x)$ are simultaneously eigenstates of the
$\mathcal{PT}$ operator: $\mathcal{PT}\phi_n(x)=\lambda_n\phi_n(x)$. As we
argued in Sec.~\ref{sec1}, $\lambda_n$ can be absorbed into $\phi_n(x)$ so that
$\mathcal{PT}\phi_n(x)=\phi_n^*(-x)=\phi_n(x)$ [see (\ref{eq8})].

\begin{figure}[t!]
\vspace{1.9in}
\includegraphics{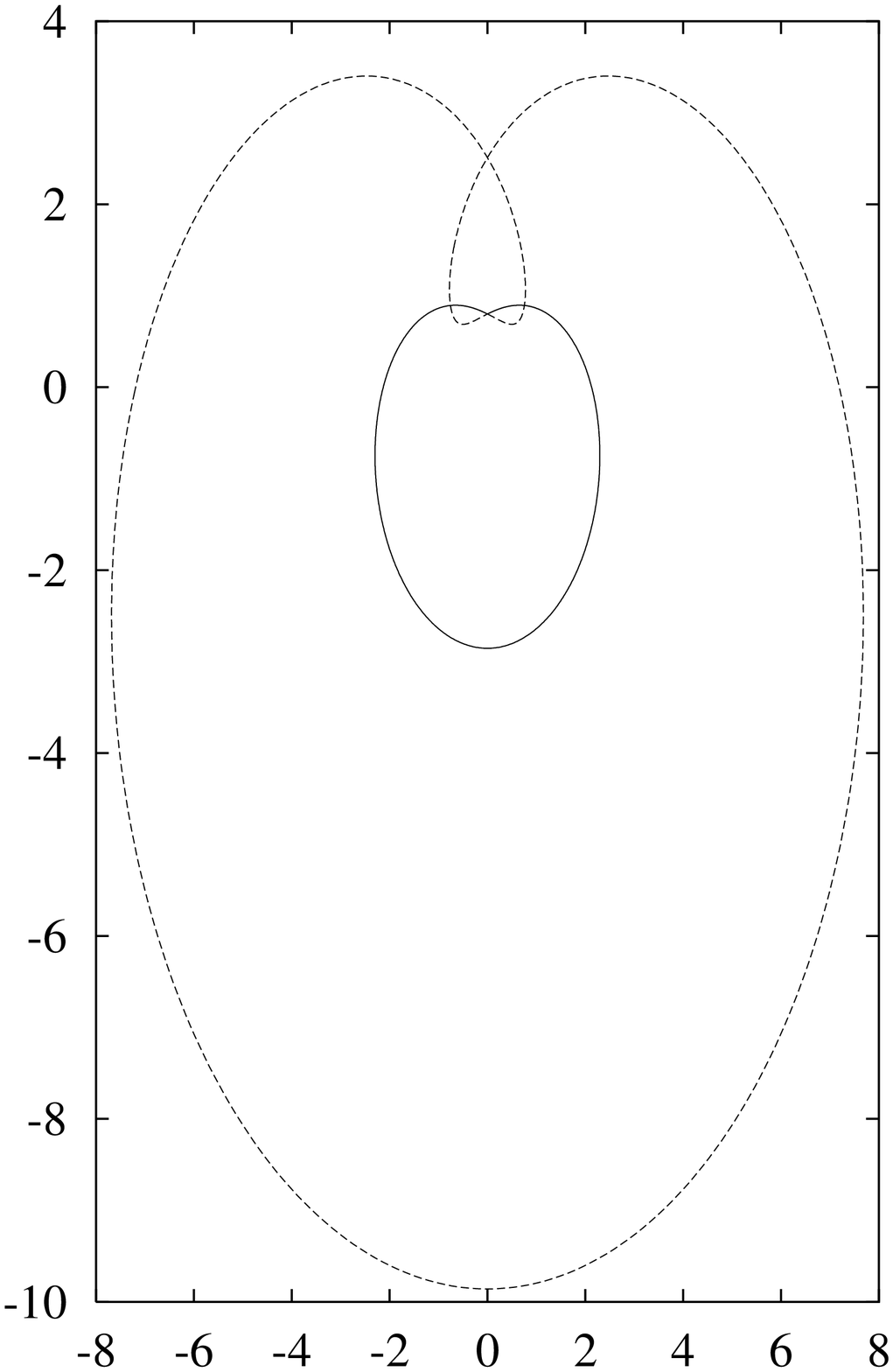}
\label{figpeter7a}
\end{figure}
\begin{figure}[t!]
\includegraphics{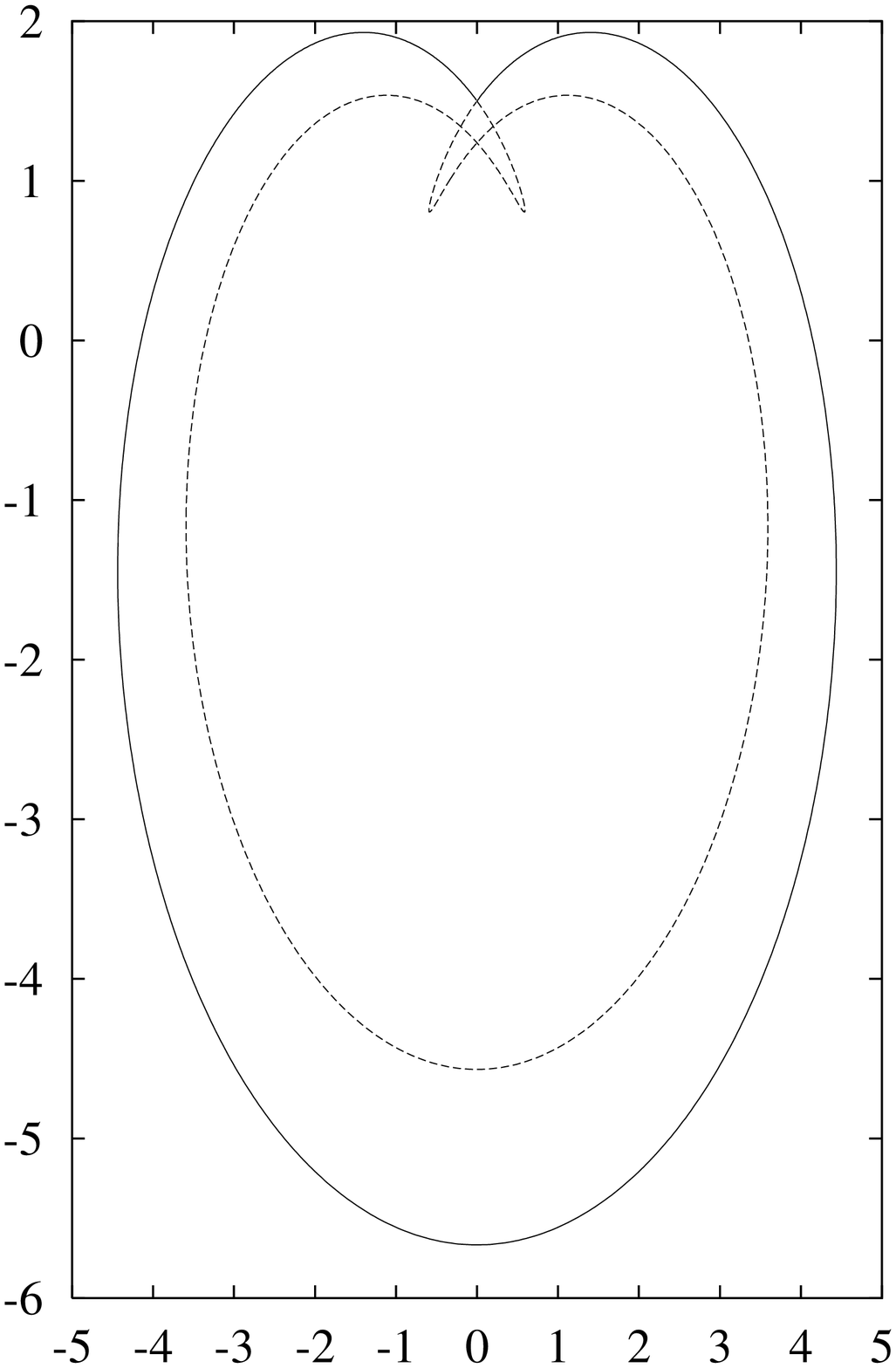}
\label{figpeter7b}
\end{figure}

The eigenstates of a conventional Hermitian Hamiltonian are complete. There is
strong evidence that the eigenfunctions $\phi_n(x)$ for the $\mathcal{P
T}$-symmetric Hamiltonian (\ref{eq6}) are also complete. The coordinate-space
statement of completeness is
\begin{equation}
\textstyle{\sum_{n=0}^\infty}(-1)^n\phi_n(x)\phi_n(y)=\delta(x-y)\qquad(x,y~{\rm
real}).
\label{eq15}
\end{equation}
This nontrivial result has been verified numerically to extremely high accuracy
(twenty decimal places) \cite{rf15,rf16}. The unusual factor of $(-1)^n$ in this
sum does not appear in conventional quantum mechanics. This factor is explained 
in the following discussion.

We must now try to find the inner product associated with our $\mathcal{P
T}$-symmetric Hamiltonian and it is here that we can see the difficulty
connected with its non-Hermiticity. In conventional Hermitian quantum mechanics 
the Hilbert space inner product is specified even before we begin to look for
the eigenstates of $H$. For our non-Hermitian Hamiltonian we must try to guess
the inner product. A reasonable guess for the inner product of two functions
$f(x)$ and $g(x)$ might be
\begin{equation}
(f,g)\equiv\textstyle{\int}dx\,[\mathcal{PT}f(x)]g(x),
\label{eq16}
\end{equation}
where $\mathcal{PT}f(x)=[f(-x)]^*$ and the path of integration in the
complex-$x$ plane follows the contour described in Sec.~\ref{sec2}. The apparent
advantage of this choice for the inner product is that the associated $\mathcal{
PT}$ norm $(f,f)$ is independent of the overall phase of $f(x)$ and is conserved
in time. With respect to this inner product the eigenfunctions $\phi_m(x)$ and
$\phi_n(x)$ of $H$ in (\ref{eq6}) are orthogonal for $n\neq m$. However, when $m
=n$ we see that the $\mathcal{PT}$ norms of the eigenfunctions are {\it not
positive}:
\begin{equation}
(\phi_m,\phi_n)=(-1)^n\delta_{mn}.
\label{eq17}
\end{equation}
This result is apparently true for all values of $N$ in (\ref{eq6}) and it has
been verified numerically to extremely high precision. Because the norms of the
eigenfunctions alternate in sign, the metric associated with the $\mathcal{PT}$
inner product $(\cdot,\cdot)$ is indefinite. This sign alternation is a {\it
generic} feature of the $\mathcal{PT}$ inner product. Extensive numerical
calculations verify that (\ref{eq17}) holds for all $N\geq2$.

\begin{figure}[t!]
\vspace{1.9in}
\includegraphics{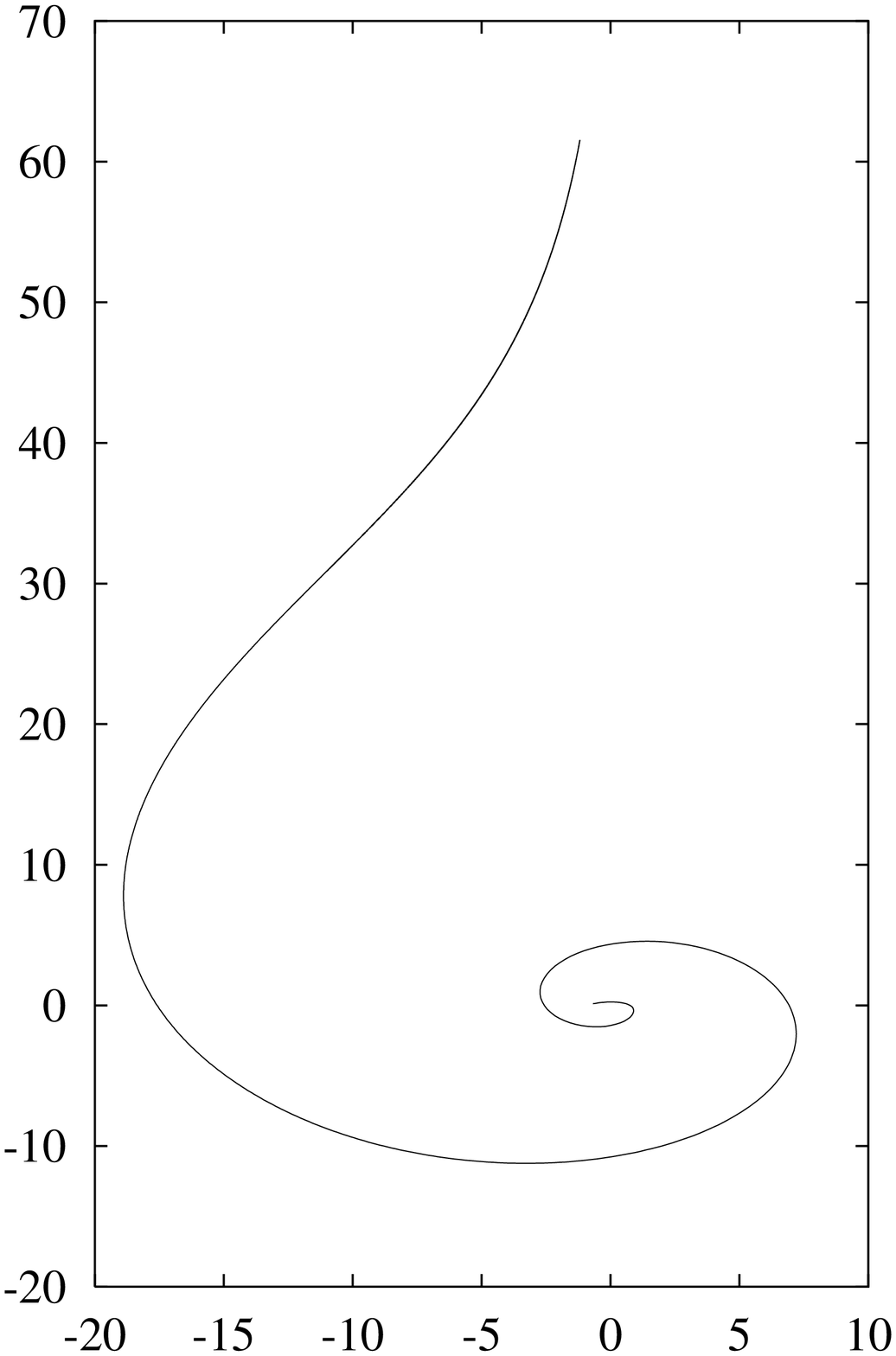}
\label{figpeter9a}
\end{figure}
\begin{figure}[t!]
\includegraphics{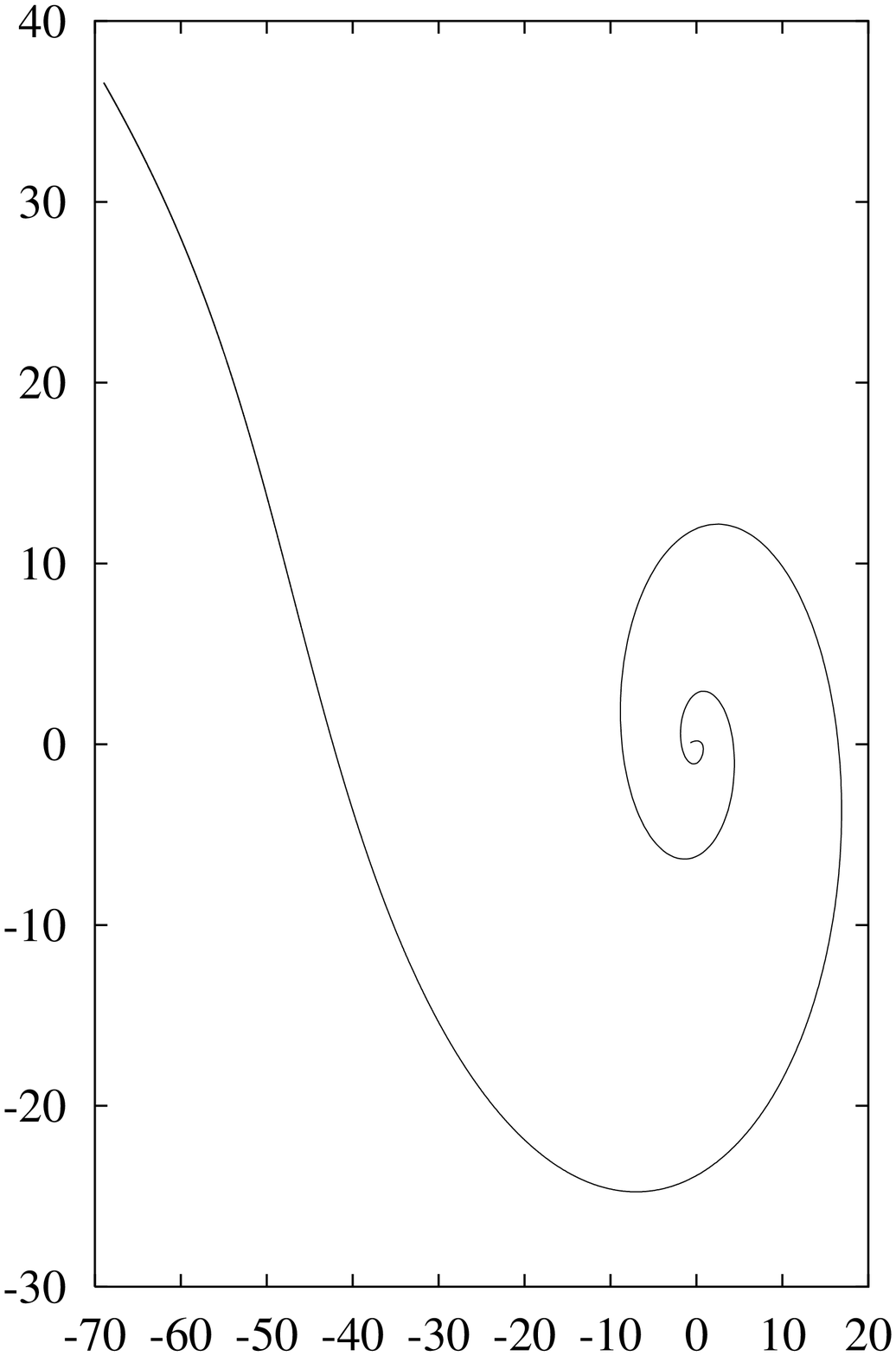}
\label{figpeter9b}
\end{figure}

Despite the existence of a nonpositive inner product, we can still do some of
the analysis that one would normally perform for a conventional Schr\"odinger
equation $H\phi_n=E_n\phi_n$. For example, we can use the inner product formula
(\ref{eq17}) to verify that (\ref{eq15}) is the representation of the unity
operator by showing that $\int dy\,\delta(x-y)\delta(y-z)=\delta(x-z)$. We can
use completeness to reconstruct the parity operator $\mathcal{P}$ in terms of
the eigenstates. The parity operator in position space is
\begin{equation}
\mathcal{P}(x,y)=\textstyle{\sum_{n=0}^\infty}(-1)^n\phi_n(x)\phi_n(-y)=\delta(
x+y).
\label{eq18}
\end{equation}
By virtue of (\ref{eq17}) the square of the parity operator is unity: $\mathcal{
P}^2=1$. We can also reconstruct $H$ in coordinate space: $H(x,y)=\sum_n(-1)^n
E_n\phi_n(x)\phi_n(y)$. Using (\ref{eq15}) - (\ref{eq17}) we can see that this
Hamiltonian satisfies $H\phi_n(x)=E_n\phi_n(x)$.

We now address the question of whether a $\mathcal{PT}$-symmetric Hamiltonian
defines a physically viable quantum mechanics. The difficulty with formulating a
$\mathcal{PT}$-symmetric quantum theory is that the vector space of quantum
states is spanned by the energy eigenstates, of which half have norm $+1$ and
half have norm $-1$. In quantum theory the norms of states carry a probabilistic
interpretation, so the indefinite metric (\ref{eq17}) is unacceptable.

The situation here in which half of the energy eigenstates have positive norm
and half have negative norm is analogous to the problem that Dirac encountered
in formulating the spinor wave equation in relativistic quantum theory
\cite{rf17}. Following Dirac, we attack the problem of an indefinite norm by
finding an interpretation of the negative-norm states. We claim that in {\it
any} theory having an unbroken $\mathcal{PT}$ symmetry there exists a symmetry
of the Hamiltonian connected with the fact that there are equal numbers of
positive- and negative-norm states. To describe this symmetry we construct the
aforementioned linear operator $\mathcal{C}$ in position space as a sum over
the eigenstates of the Hamiltonian \cite{rf14}:
\begin{equation}
\mathcal{C}(x,y)=\textstyle{\sum_{n=0}^\infty}\phi_n(x)\phi_n(y).
\label{eq21}
\end{equation}

The properties of this new operator $\mathcal{C}$ resemble those of the charge
conjugation operator in quantum field theory. For example, we can use
(\ref{eq15}) - (\ref{eq17}) to verify that the square of $\mathcal{C}$ is unity
($\mathcal{C}^2=1$): $\int dy\,\mathcal{C}(x,y)\mathcal{C}(y,z)=\delta(x-z)$.
Thus, the eigenvalues of $\mathcal{C}$ are $\pm1$. Also, $\mathcal{C}$ commutes
with the Hamiltonian $H$. Therefore, since $\mathcal{C}$ is linear, the
eigenstates of $H$ have definite values of $\mathcal{C}$. Specifically, if the
energy eigenstates satisfy (\ref{eq17}), then we have
$$\mathcal{C}\phi_n(x)=\textstyle{\int}dy\,\mathcal{C}(x,y)\phi_n(y)=\textstyle{
\sum_{m=0}^\infty}\phi_m(x)\textstyle{\int}dy\,\phi_m(y)\phi_n(y)=(-1)^n\phi_n(
x).$$
Thus, $\mathcal{C}$ represents the measurement of the sign of the $\mathcal{PT}$
norm of an eigenstate in (\ref{eq17}).

The operators $\mathcal{P}$ and $\mathcal{C}$ are distinct square roots of the
unity operator $\delta(x-y)$. That is, $\mathcal{P}^2=\mathcal{C}^2=1$, but
$\mathcal{P}\neq\mathcal{C}$. Indeed, $\mathcal{P}$ is real, while $\mathcal{C}$
is complex. The parity operator in coordinate space is explicitly real
$\mathcal{P}(x,y)=\delta(x+y)$, while the operator $\mathcal{C}(x,y)$ is complex
because it is a sum of products of complex functions, as we see in (\ref{eq21}).
The two operators $\mathcal{P}$ and $\mathcal{C}$ do not commute. However,
$\mathcal{C}$ {\it does} commute with $\mathcal{PT}$.

Finally, having obtained the operator $\mathcal{C}$ we define the new inner
product structure given in (\ref{eq14}). This inner product has a {\it positive
definite} norm. Like the $\mathcal{PT}$ inner product (\ref{eq16}) this
new inner product is phase independent. Also, it is conserved in time because
the time evolution operator (just as in ordinary quantum mechanics) is $e^{iH
t}$. The fact that $H$ commutes with $\mathcal{PT}$ and with $\mathcal{CPT}$
implies that both inner products, (\ref{eq16}) and (\ref{eq14}), remain time
independent as the states evolve. However, unlike (\ref{eq16}), the inner
product (\ref{eq14}) is positive definite because $\mathcal{C}$ contributes $-1$
when it acts on states with negative $\mathcal{PT}$ norm. In terms of the
$\mathcal{CPT}$ conjugate, the completeness condition (\ref{eq15}) reads
\begin{equation}
\textstyle{\sum_{n=0}^\infty}\phi_n(x)[\mathcal{CPT}\phi_n(y)]=\delta(x-y).
\label{eq24}
\end{equation}

To review, in the mathematical formulation of a conventional quantum theory the
Hilbert space of physical states is specified first. The inner product in this
vector space is defined with respect to ordinary Dirac Hermitian conjugation
(complex conjugate and transpose). The Hamiltonian is then chosen and the
eigenvectors and eigenvalues of the Hamiltonian are determined. In contrast, the
inner product for a quantum theory defined by a non-Hermitian $\mathcal{P
T}$-symmetric Hamiltonian depends on the Hamiltonian itself and is thus
determined {\it dynamically}. One can view this new kind of quantum theory as a
``bootstrap'' theory because one must solve for the eigenstates of $H$ before
knowing what the Hilbert space and the associated inner product of the theory
are. The Hilbert space and the $\mathcal{CPT}$ inner product (\ref{eq14}) are
then determined by these eigenstates via (\ref{eq21}).

The operator $\mathcal{C}$ does not exist as a distinct entity in ordinary
Hermitian quantum mechanics. Indeed, if we allow the parameter $N$ in
(\ref{eq6}) to tend to 2, the operator $\mathcal{C}$ in this limit becomes
identical to $\mathcal{P}$. Thus, in this limit the $\mathcal{CPT}$ operator
becomes $\mathcal{T}$, which is just complex conjugation. As a consequence, the
inner product (\ref{eq14}) defined with respect to the $\mathcal{CPT}$
conjugation reduces to the complex conjugate inner product of conventional
quantum mechanics when $N\to2$. Similarly, in this limit (\ref{eq24}) reduces to
the usual statement of completeness $\sum_n\phi_n(x)\phi_n^*(y)=\delta(x-y)$.

The $\mathcal{CPT}$ inner-product (\ref{eq14}) is independent of the choice of
integration contour ${\rm C}$ as long as ${\rm C}$ lies inside the asymptotic
wedges associated with the boundary conditions for the eigenvalue problem
(\ref{eq9}). In ordinary quantum mechanics, where the positive-definite inner
product has the form $\int dx\,f^*(x)g(x)$, the integral must be taken along the
real axis and the path of the integration cannot be deformed into the complex
plane because the integrand is not analytic \cite{foot2}. The $\mathcal{PT}$
inner product (\ref{eq16}) shares with (\ref{eq14}) the advantage of analyticity
and path independence, but it suffers from nonpositivity. It is surprising that
we can construct a positive-definite metric by using $\mathcal{CPT}$ conjugation
without disturbing the path independence of the inner-product integral.

We can now explain why $\mathcal{PT}$-symmetric theories are unitary. Time
evolution is expressed by the operator $e^{-iHt}$, whether the theory is
determined by a $\mathcal{PT}$-symmetric Hamiltonian or just an ordinary
Hermitian Hamiltonian. To establish unitarity we must show that as a state
vector evolves, its norm does not change in time. If $\psi_0(x)$ is a prescribed
initial wave function belonging to the Hilbert space spanned by the energy
eigenstates, then it evolves into the state $\psi_t(x)$ at time $t$ according to
$\psi_t(x)=e^{-iHt}\psi_0(x)$. With respect to the $\mathcal{CPT}$ inner product
defined in (\ref{eq14}), the norm of the vector $\psi_t(x)$ does not change in
time, $\langle\psi_t|\psi_t\rangle=\langle\psi_0|\psi_0\rangle$, because the
Hamiltonian $H$ commutes with the $\mathcal{CPT}$ operator.

\section{Illustrative Example: A $2\times2$ Matrix Hamiltonian}
\label{sec5}

The $2\times2$ matrix Hamiltonian
\begin{equation}
H=\left(\begin{array}{cc} re^{i\theta} & s \cr s & re^{-i\theta}
\end{array}\right)
\label{eq25}
\end{equation}
where the three parameters $r$, $s$, and $\theta$ are real, illustrates the
above results on $\mathcal{PT}$-symmetric quantum mechanics. This Hamiltonian is
not Hermitian, but it is $\mathcal{PT}$ symmetric, where the parity operator is
$\mathcal{P}=\left({0\atop1}\,{1\atop0}\right)$ and $\mathcal{T}$ performs
complex conjugation \cite{rf20}.

There are two parametric regions for this Hamiltonian. When $s^2<r^2\sin^2
\theta$, the energy eigenvalues form a complex conjugate pair. This is the
region of broken $\mathcal{PT}$ symmetry. On the other hand, when $s^2\geq r^2
\sin^2\theta$, then the eigenvalues $\varepsilon_\pm=r\cos\theta\pm(s^2-r^2
\sin^2\theta)^{1/2}$ are real. This is the region of unbroken $\mathcal{PT}$
symmetry. In the unbroken region the simultaneous eigenstates of the operators
$H$ and $\mathcal{PT}$ are
\begin{equation}
|\varepsilon_+\rangle=\frac{1}{\sqrt{2\cos\alpha}}\left(\begin{array}{c}
e^{i\alpha/2}\cr e^{-i\alpha/2}\end{array}\right)\quad{\rm and}\quad
|\varepsilon_-\rangle=\frac{i}{\sqrt{2\cos\alpha}}\left(\begin{array}{c}
e^{-i\alpha/2}\cr-e^{i\alpha/2}\end{array}\right),
\label{eq27}
\end{equation}
where we set $\sin\alpha =(r/s)\,\sin\theta$. The $\mathcal{PT}$ inner product
gives $(\varepsilon_{\pm},\varepsilon_{\pm})=\pm1$ and $(\varepsilon_{\pm},
\varepsilon_{\mp})=0$, where $(u,v)=(\mathcal{PT}u)\cdot v$. Therefore, with
respect to the $\mathcal{PT}$ inner product, the resulting vector space spanned
by the energy eigenstates has a metric of signature $(+,-)$. The condition $s^2>
r^2\sin^2\theta$ ensures that the $\mathcal{PT}$ symmetry is not broken. If this
condition is violated, the states (\ref{eq27}) are no longer eigenstates of
$\mathcal{PT}$ because $\alpha$ becomes imaginary. When $\mathcal{PT}$ symmetry
is broken, the $\mathcal{PT}$ norm of the energy eigenstate vanishes.

Next, we construct the operator $\mathcal{C}$ using (\ref{eq21}):
\begin{equation}
\mathcal{C}=\frac{1}{\cos\alpha}\left(\begin{array}{cc} i\sin\alpha &
1 \cr 1 & -i\sin\alpha \end{array}\right).
\label{eq28}
\end{equation}
Note that $\mathcal{C}$ is distinct from $H$ and $\mathcal{P}$ and it has the
key property that $\mathcal{C}|\varepsilon_{\pm}\rangle=\pm|\varepsilon_{\pm}
\rangle$. The operator $\mathcal{C}$ commutes with $H$ and satisfies
$\mathcal{C}^2=1$. The eigenvalues of $\mathcal{C}$ are precisely the signs of
the $\mathcal{PT}$ norms of the corresponding eigenstates. Using the operator
$\mathcal{C}$ we construct the new inner product structure $\langle u|v\rangle=
(\mathcal{CPT}u)\cdot v$. This inner product is positive definite because
$\langle\varepsilon_{\pm}| \varepsilon_{\pm}\rangle=1$. Thus, the
two-dimensional Hilbert space spanned by $|\varepsilon_\pm\rangle$, with inner
product $\langle\cdot|\cdot\rangle$, has signature $(+,+)$.

Finally, we show that the $\mathcal{CPT}$ norm of any vector is positive. For
the arbitrary vector $\psi=\left({a\atop b}\right)$, where $a$ and $b$ are any
complex numbers, we see that
$$\mathcal{T}\psi=\left(a^*\atop b^*\right),\quad\mathcal{PT}\psi=\left(b^*\atop
a^*\right),~{\rm and}\quad\mathcal{CPT}\psi={1\over\cos\alpha}\,\left(a^*+ib^*
\sin\alpha\atop b^*-ia^*\sin\alpha\right).$$
Thus, $\langle\psi|\psi\rangle=(\mathcal{CPT}\psi)\cdot\psi=\frac{1}{\cos\alpha}
[a^*a+b^*b+i(b^*b-a^*a)\sin\alpha]$. Now let $a=x+iy$ and $b=u+iv$, where $x$,
$y$, $u$, and $v$ are real. Then
\begin{equation}
\langle\psi|\psi\rangle=\frac{1}{\cos\alpha}\left(x^2+v^2+2xv\sin\alpha
+y^2+u^2-2yu\sin\alpha\right), 
\label{eq29}
\end{equation}
which is explicitly positive and vanishes only if $x=y=u=v=0$.

Since $\langle u|$ denotes the $\mathcal{CPT}$-conjugate of $|u\rangle$, the
completeness condition reads
\begin{equation}
|\varepsilon_+\rangle\langle\varepsilon_+|+|\varepsilon_-\rangle\langle
\varepsilon_-|=\left(\begin{array}{cc} 1 & 0 \cr 0 & 1\end{array}\right). 
\label{eq30}
\end{equation}
Furthermore, using the $\mathcal{CPT}$ conjugate $\langle\varepsilon_\pm|$, we
get $\mathcal{C}$ as $\mathcal{C}=|\varepsilon_+\rangle\langle\varepsilon_+|-|
\varepsilon_-\rangle\langle\varepsilon_-|$.

If we set $\theta=0$ in this two-state system, the Hamiltonian (\ref{eq25})
becomes Hermitian. However, $\mathcal{C}$ then reduces to the parity operator
$\mathcal{P}$. As a consequence, $\mathcal{CPT}$ invariance reduces to the
standard condition of Hermiticity for a symmetric matrix; namely, that $H=H^*$.
This is why the hidden symmetry $\mathcal{C}$ was not noticed previously. The
operator $\mathcal{C}$ emerges only when we extend a real symmetric Hamiltonian 
into the complex domain.

\section{Observables in $\mathcal{PT}$-Symmetric Quantum Mechanics}
\label{sec6}
How do we represent an observable in $\mathcal{PT}$-symmetric quantum mechanics?
Recall that in ordinary quantum mechanics the condition for a linear operator
$A$ to be an observable is that $A=A^\dagger$. This condition guarantees that
the expectation value of $A$ in a state is real. Because operators in the
Heisenberg picture evolve in time according to $A(t)=e^{iHt}A(0)e^{-iHt}$,
this Hermiticity condition is maintained in time. In $\mathcal{PT}$-symmetric
quantum mechanics the equivalent condition is that at time $t=0$ the operator
$A$ must obey the condition $A^{\rm T}=\mathcal{CPT}A\,\mathcal{CPT}$, where
$A^{\rm T}$ is the {\it transpose} of $A$. If this condition holds at $t=0$,
then it will continue to hold for all time because we have assumed that $H$ is
symmetric ($H=H^{\rm T}$). This condition also guarantees that the expectation
value of $A$ in any state is real.

The operator $\mathcal{C}$ itself satisfies this requirement, so it is an
observable. Also the Hamiltonian is an observable. However, the $x$ and $p$
operators are not observables. Indeed, the expectation value of $x$ in the
ground state is a negative imaginary number. Thus, there is no position operator
in $\mathcal{PT}$-symmetric quantum mechanics. In this sense $\mathcal{P
T}$-symmetric quantum mechanics is similar to fermionic quantum field theories.
In such theories the fermion field corresponds to the $x$ operator. The fermion
field is complex and does not have a classical limit. One cannot measure the
position of an electron; one can only measure the position of the {\it charge}
or of the {\it energy} of the electron!

One can see why the expectation of the $x$ operator is a negative imaginary
number by examining Fig.~\ref{figpeter2}. Note that the classical trajectories
have left-right ($\mathcal{PT}$) symmetry, but not up-down symmetry. Also, the
classical paths favor the lower-half complex-$x$ plane. Thus, the average
classical position is a negative imaginary number. Just as the classical
particle moves about in the complex plane, the quantum probability current
flows about in the complex plane. It may be that the correct interpretation
is to view $\mathcal{PT}$-symmetric quantum mechanics as describing the
interaction of extended, rather than pointlike objects.

\section{Calculation of the $\mathcal{C}$ Operator}
\label{sec7}

The distinguishing feature of $\mathcal{PT}$-symmetric quantum mechanics is the
$\mathcal{C}$ operator. The discovery of $\mathcal{C}$ in Ref.~\cite{rf14}
raises the question of how to evaluate the formal sum in (\ref{eq21}) that
represents $\mathcal{C}$. In ordinary Hermitian quantum mechanics there is no
such operator. Only a {\it non-Hermitian} $\mathcal{PT}$-symmetric Hamiltonian
possesses a $\mathcal{C}$ operator distinct from the parity operator $\mathcal{P
}$. Indeed, if we were to evaluate (\ref{eq21}) for a $\mathcal{PT}$-symmetric
Hamiltonian that is also Hermitian, the result would be $\mathcal{P}$, which in
coordinate space is $\delta(x+y)$ [see (\ref{eq18})].

Calculating $\mathcal{C}$ by direct evaluation of the sum in (\ref{eq21}) is
not easy in quantum mechanics because it is necessary to calculate all the
eigenfunctions $\phi_n(x)$ of $H$. Such a procedure cannot be used in quantum
field theory because there is no simple analog of the Schr\"odinger eigenvalue
differential equation and its associated coordinate-space eigenfunctions.

Fortunately, there is an easy way to calculate the $\mathcal{C}$ operator, and
the procedure circumvents the difficult problem of evaluating the sum in
(\ref{eq21}). As a result the technique readily generalizes from quantum
mechanics to quantum field theory. In this section we use this technique to
calculate $\mathcal{C}$ for the $\mathcal{PT}$-symmetric Hamiltonian \cite{rx}
\begin{eqnarray}
H=\half{\hat p}^2+\half{\hat x}^2+i\epsilon{\hat x}^3.
\label{eee10}
\end{eqnarray}
We will show how to calculate $\mathcal{C}$ perturbatively to high order in
powers of $\epsilon$ for this cubic Hamiltonian. Calculating $\mathcal{C}$ for
other kinds of interactions is a bit more difficult and requires the use of
semiclassical approximations \cite{rf20.5}.

Our calculation of $\mathcal{C}$ makes use of its three crucial properties.
First, $\mathcal{C}$ commutes with the space-time reflection operator $\mathcal{
PT}$,
\begin{eqnarray}
[\mathcal{C},\mathcal{PT}]=0,
\label{eee2}
\end{eqnarray}
although $\mathcal{C}$ does not commute with $\mathcal{P}$ or $\mathcal{T}$
separately. Second, the square of $\mathcal{C}$ is the identity,
\begin{eqnarray}
\mathcal{C}^2={\bf 1},
\label{eee3}
\end{eqnarray}
which allows us to interpret $\mathcal{C}$ as a reflection operator. Third,
$\mathcal{C}$ commutes with $H$,
\begin{eqnarray}
[\mathcal{C},H]=0,
\label{eee4}
\end{eqnarray}
and thus is time independent. To summarize, $\mathcal{C}$ is a time-independent
$\mathcal{PT}$-symmetric reflection operator.

The procedure for calculating $\mathcal{C}$ begins by introducing a general
operator representation for $\mathcal{C}$ of the form
\cite{rf18}
\begin{eqnarray}
\mathcal{C}=e^{Q({\hat x},{\hat p})}\mathcal{P},
\label{eee14}
\end{eqnarray}
where $\mathcal{P}$ is the parity operator and $Q({\hat x},{\hat p})$ is a real
function of the dynamical variables ${\hat x}$ and ${\hat p}$. This
representation conveniently incorporates the three requirements (\ref{eee2}) -
(\ref{eee4}). 

The representation $\mathcal{C}=e^Q\mathcal{P}$ is general. Let us illustrate
this simple representation for $\mathcal{C}$ in two elementary cases: First,
consider the shifted harmonic oscillator $H=\half{\hat p}^2+\half{\hat x}^2+i
\epsilon{\hat x}$. This Hamiltonian has an unbroken $\mathcal{PT}$ symmetry for
all real $\epsilon$. Its eigenvalues $E_n=n+\half+\half\epsilon^2$ are all real.
The $\mathcal{C}$ operator for this theory is given exactly by $\mathcal{C}=e^Q
\mathcal{P}$, where $Q=-\epsilon{\hat p}$. Note that in the limit $\epsilon\to
0$, where the Hamiltonian becomes Hermitian, $\mathcal{C}$ becomes identical
with $\mathcal{P}$.

As a second example, consider the non-Hermitian $2\times2$ matrix Hamiltonian
(\ref{eq25}). The $\mathcal{C}$ operator in (\ref{eq28}) can be easily
rewritten in the form $\mathcal{C}=e^Q\mathcal{P}$, where $Q=\half\sigma_2\ln
\left(\frac{1-\sin\alpha}{1+\sin\alpha}\right)$. Here, $\sigma_2=\left({0~-i
\atop\!\!i~\,\,0}\right)$. Again, observe that in the limit $\theta\to0$, where
the Hamiltonian becomes Hermitian, the $\mathcal{C}$ operator becomes identical
with $\mathcal{P}$.

We will now calculate $\mathcal{C}$ directly from its operator representation
(\ref{eee14}) and we will show that $Q({\hat x},{\hat p})$ can be found by
solving elementary operator equations. To find the operator equations satisfied
by $Q$ we substitute $\mathcal{C}=e^Q\mathcal{P}$ into the three equations
(\ref{eee2}) - (\ref{eee4}) in turn.

First, we substitute (\ref{eee14}) into the condition (\ref{eee2}) to obtain
$$e^{Q({\hat x},{\hat p})}=\mathcal{PT}e^{Q({\hat x},{\hat p})}\mathcal{PT}=
e^{Q(-{\hat x},{\hat p})},$$
from which we conclude that $Q({\hat x},{\hat p})$ is an {\it even} function of
${\hat x}$. Second, we substitute (\ref{eee14}) into the condition (\ref{eee3})
and find that
$$e^{Q({\hat x},{\hat p})}\mathcal{P}e^{Q({\hat x},{\hat p})}\mathcal{P}=e^{Q(
{\hat x},{\hat p})}e^{Q(-{\hat x},-{\hat p})}=1,$$
which implies that $Q({\hat x},{\hat p})=-Q(-{\hat x},-{\hat p})$. Since we
already know that $Q({\hat x},{\hat p})$ is an even function of ${\hat x}$, we
conclude that it is also an {\it odd} function of ${\hat p}$.

The remaining condition (\ref{eee4}) to be imposed is that the operator
$\mathcal{C}$ commutes with $H$. Substituting $\mathcal{C}=e^{Q({\hat x},{\hat
p})}\mathcal{P}$ into (\ref{eee4}), we get $e^{Q({\hat x},{\hat p})}[\mathcal{P}
,H]+[e^{Q({\hat x},{\hat p})},H]\mathcal{P}=0$. We can express the Hamiltonian
$H$ in (\ref{eee10}) in the form $H=H_0+\epsilon H_1$, where $H_0$ is the
harmonic oscillator Hamiltonian $H_0=\half{\hat p}^2+\half{\hat x}^2$, which
commutes with the parity operator $\mathcal{P}$, and $H_1=i{\hat x}^3$, which
{\it anticommutes} with $\mathcal{P}$. The above condition becomes
\begin{eqnarray}
2\epsilon e^{Q({\hat x},{\hat p})}H_1=[e^{Q({\hat x},{\hat p})},H].
\label{eee29}
\end{eqnarray}

The operator $Q({\hat x},{\hat p})$ may be expanded as a series in odd powers of
$\epsilon$:
\begin{eqnarray}
Q({\hat x},{\hat p})=\epsilon Q_1({\hat x},{\hat p})+\epsilon^3 Q_3({\hat x},
{\hat p})+\epsilon^5Q_5({\hat x},{\hat p})+\cdots\,.
\label{eee30}
\end{eqnarray}
Substituting the expansion in (\ref{eee30}) into the exponential $e^{Q({\hat x},
{\hat p})}$, we get after some algebra a sequence of equations that can be
solved systematically for the operator-valued functions $Q_n({\hat x},{\hat p})$
$(n=1,3,5,\ldots)$ subject to the symmetry constraints that ensure the
conditions (\ref{eee2}) and (\ref{eee3}). The first three of these equations are
\begin{eqnarray}
\left[H_0,Q_1\right] &=& -2H_1,\nonumber\\
\left[H_0,Q_3\right] &=& -{\textstyle\frac{1}{6}}[Q_1,[Q_1,H_1]],\nonumber\\
\left[H_0,Q_5\right] &=& {\textstyle\frac{1}{360}}[Q_1,[Q_1,[Q_1,[Q_1,H_1]]]]
-{\textstyle\frac{1}{6}}\left([Q_1,[Q_3,H_1]]+[Q_3,[Q_1,H_1]]\right).
\label{eee34}
\end{eqnarray}

Let us solve these equations for the Hamiltonian in (\ref{eee10}), for which
$H_0=\half{\hat p}^2+\half{\hat x}^2$ and $H_1=i{\hat x}^3$. The procedure is to
substitute the most general polynomial form for $Q_n$ using arbitrary
coefficients and then to solve for these coefficients. For example, to solve the
first of the equations in (\ref{eee34}), $\left[H_0,Q_1\right]=-2i{\hat x}^3$,
we take as an {\it ansatz} for $Q_1$ the most general Hermitian cubic polynomial
that is even in ${\hat x}$ and odd in $p$:
\begin{eqnarray}
Q_1({\hat x},{\hat p})=M{\hat p}^3+N{\hat x}{\hat p}{\hat x},
\label{eee35}
\end{eqnarray}
where $M$ and $N$ are undetermined coefficients. The operator equation for $Q_1$
is satisfied if $M=-{\textstyle\frac{4}{3}}$ and $N=-2$.

It is straightforward, though somewhat tedious, to continue this process. In
order to present the solutions for $Q_n({\hat x},{\hat p})$ ($n>1$), it is
convenient to introduce the following notation: Let $S_{m,n}$ represent the {\it
totally symmetrized} sum over all terms containing $m$ factors of ${\hat p}$ and
$n$ factors of ${\hat x}$. For example, $S_{0,0}=1$, $S_{0,3}={\hat x}^3$, $S_{1
,1}=\half\left({\hat x}{\hat p}+{\hat p}{\hat x}\right)$, $S_{1,2}={\textstyle
\frac{1}{3}}\left({\hat x}^2{\hat p}+{\hat x}{\hat p}{\hat x}+{\hat p}{\hat
x}^2\right)$, and so on. The properties of the operators $S_{m,n}$ are
summarized in Ref.~\cite{rf23}

In terms of the symmetrized operators $S_{m,n}$ the first three functions
$Q_{2n+1}$ are
\begin{eqnarray}
Q_1 \!&=&\! -{\textstyle\frac{4}{3}}{\hat p}^3-2S_{1,2},\nonumber\\
Q_3 \!&=&\! {\textstyle\frac{128}{15}}{\hat p}^5+{\textstyle\frac{40}{3}}
S_{3,2}+8S_{1,4}-12{\hat p},\nonumber\\
Q_5 \!&=&\! -{\textstyle\frac{320}{3}}{\hat p}^7-{\textstyle\frac{544}{3}}
S_{5,2}-{\textstyle\frac{512}{3}}S_{3,4}
-64S_{1,6}+{\textstyle\frac{24\,736}{45}}{\hat p}^3
+ {\textstyle\frac{6\,368}{15}}S_{1,2}.
\label{eee39}
\end{eqnarray}
This completes the calculation of $\mathcal{C}$. Together, (\ref{eee14}),
(\ref{eee30}), and (\ref{eee39}) represent an explicit perturbative expansion of
$\mathcal{C}$ in terms of the operators ${\hat x}$ and ${\hat p}$, correct to
order $\epsilon^6$.

To summarize, using the {\it ansatz} (\ref{eee14}) we can calculate the
$\mathcal{C}$ operator to very high order in perturbation theory. We are able to
perform this calculation because this {\it ansatz} obviates the necessity of
calculating the wave functions $\phi_n(x)$.

\section{Quantum Mechanics in the Complex Plane}
\label{sec8}
We have seen in Sec.~\ref{sec3} that the classical motion of particles described
by $\mathcal{PT}$-symmetric Hamiltonians is not confined to the real-$x$ axis;
the classical paths of such particles lie in the complex-$x$ plane. Analogously,
the new kinds of quantum theories discussed in this paper may also be viewed as
extensions of ordinary quantum mechanics into the complex domain. This is so
because, as we saw in Sec.~\ref{sec2}, the Schr\"odinger equation eigenvalue
problem and the corresponding boundary conditions are posed in the complex-$x$
plane.

The idea of extending a Hermitian Hamiltonian into the complex plane was first
discussed by Dyson, who argued heuristically that perturbation theory for
quantum electrodynamics diverges \cite{DY}. Dyson's argument involves rotating
the electric charge $e$ into the complex plane $e\to ie$. Applied to the
anharmonic oscillator Hamiltonian
\begin{equation}
H=\half{\hat p}^2+\half{\hat x}^2+\textstyle{\frac{1}{4}}g{\hat x}^4\qquad(g>0),
\label{eq101}
\end{equation}
Dyson's argument would go as follows: Rotate the coupling $g$ into the
complex-$g$ plane to $-g$. Then the potential is no longer bounded below, so the
resulting theory has no ground state. Thus, the energies $E_n(g)$ are singular
at $g=0$ and the perturbation series for $E_n(g)$, which are series in powers of
$g$, must therefore have a zero radius of convergence and must diverge for all
$g\neq0$. These perturbation series do indeed diverge, but there is flaw in
Dyson's argument, and understanding this flaw is necessary to understand how a
non-Hermitian $\mathcal{PT}$-symmetric Hamiltonian can have a positive real
spectrum.

The flaw in Dyson's argument is simply that the eigenvalues of the Hamiltonian
\begin{equation}
H=\half{\hat p}^2+\half{\hat x}^2-\textstyle{\frac{1}{4}}g{\hat x}^4\qquad(g>0),
\label{eq104}
\end{equation}
are undefined until the boundary conditions on the eigenfunctions are specified.
These boundary conditions depend crucially on how this Hamiltonian with negative
coupling is obtained. Dyson's way to obtain $H$ in (\ref{eq104}) would be to
substitute $g=|g|e^{i\theta}$ into (\ref{eq101}) and to rotate from $\theta=0$
to $\theta=\pi$. Under this rotation, the energies $E_n(g)$ become complex: The
$E_n(g)$ are real and positive when $g>0$ but complex when $g<0$. The $\mathcal{
PT}$-symmetric way to obtain (\ref{eq104}) is to take the limit $\delta$: $0\to
2$ of $H=\half{\hat p}^2+\half{\hat x}^2+\textstyle{\frac{1}{4}}g{\hat x}^2(i{
\hat x})^\delta$ $(g>0)$. When (\ref{eq104}) is obtained by this limiting
process, its spectrum is real, positive, and discrete.

How can the Hamiltonian (\ref{eq104}) possess two such astonishingly different
spectra? As we saw in Sec.~\ref{sec2}, the answer lies in understanding the
boundary conditions satisfied by the wave functions $\phi_n(x)$. Under Dyson's
rotation the eigenfunctions $\phi_n(x)$ vanish in the complex-$x$ plane as $|x|
\to\infty$ inside the wedges $-\pi/3<{\rm arg}\,x<0$ and $-4\pi/3<{\rm arg}\,x<
-\pi$. Under the $\mathcal{PT}$ limiting process, in which the exponent $\delta$
ranges from $0$ to $2$, $\phi_n(x)$ vanishes in the complex-$x$ plane as $|x|\to
\infty$ inside the wedges $-\pi/3<{\rm arg}\,x<0$ and $-\pi<{\rm arg} \,x<-2\pi/
3$. In the latter case the boundary conditions hold in wedges that are symmetric
with respect to the imaginary axis (see Fig.~\ref{figcmb2}); these boundary
conditions enforce the $\mathcal{PT}$ symmetry of $H$ and are responsible for
the reality of the energy spectrum.

Apart from the differences in the energy levels, there is another striking
difference between the two theories corresponding to $H$ in (\ref{eq104}). Under
Dyson's rotation the expectation value of the operator ${\hat x}$ remains zero.
This is because Dyson's rotation preserves the parity symmetry of $H$ in
(\ref{eq104}). However, under our limiting process, in which $\delta$ ranges
from $0$ to $2$, this expectation value becomes nonzero because as soon as
$\delta$ begins to increase, parity symmetry is violated (and is replaced by
$\mathcal{PT}$ symmetry.

The nonvanishing of the expectation value of ${\hat x}$ has important physical
consequences. We suggest in the next section that $\mathcal{PT}$ symmetry
may be the ideal quantum field theoretic setting to describe the dynamics of the
Higgs sector in the standard model of particle physics.

\section{Physical Applications of $\mathcal{PT}$-Symmetric Quantum Theories}
\label{sec9}

It is not known whether non-Hermitian, $\mathcal{PT}$-symmetric Hamiltonians can
be used to describe experimentally observable phenomena. However, non-Hermitian 
Hamiltonians have {\it already} been used to describe interesting interacting
systems. For example, Wu showed that the ground state of a Bose system of hard
spheres is described by a non-Hermitian Hamiltonian \cite{rf30}. Wu found that
the ground-state energy of this system is real and conjectured that all the
energy levels were real. Hollowood showed that even though the Hamiltonian of a
complex Toda lattice is non-Hermitian, the energy levels are real \cite{rf31}.
Non-Hermitian Hamiltonians of the form $H={\hat p}^2+i{\hat x}^3$ and cubic
quantum field theories arise in studies of the Lee-Yang edge singularity
\cite{rf1} and in various Reggeon field theory models \cite{rf2}. In each of
these cases the fact that a non-Hermitian Hamiltonian had a real spectrum
appeared mysterious at the time, but now the explanation is simple: In each case
the non-Hermitian Hamiltonian is $\mathcal{PT}$-symmetric. In each case the
Hamiltonian is constructed so that the position operator ${\bar x}$ or the field
operator $\phi$ is always multiplied by $i$.

An experimental signal of a complex Hamiltonian might be found in the context of
condensed matter physics. Consider the complex crystal lattice whose potential
is $V(x)=i\sin\,x$. While the Hamiltonian $H={\hat p}^2+i\sin\,{\hat x}$ is not
Hermitian, it is $\mathcal{PT}$-symmetric and all of its energy bands are {\it
real}. However, at the edge of the bands the wave function of a particle in such
a lattice is always bosonic ($2\pi$-periodic), and unlike the case of ordinary
crystal lattices, the wave function is never fermionic ($4\pi$-periodic)
\cite{rf32}. Direct observation of such a band structure would give unambiguous
evidence of a $\mathcal{PT}$-symmetric Hamiltonian.

The quartic $\mathcal{PT}$-symmetric quantum field theory that corresponds to
$H$ in (\ref{eq6}) with $N=4$ is described by the ``wrong-sign'' Hamiltonian
density
\begin{equation}
\mathcal{H}=\half\pi^2({\bf x},t)+\half[\nabla_{\!\bf x}\varphi({\bf x},t)]^2+
\half\mu^2\varphi^2({\bf x},t)-{\textstyle\frac{1}{4}}g\varphi^4({\bf x},t).
\label{eq100}
\end{equation}
This theory is remarkable because, in addition to the energy spectrum being real
and positive, the one-point Green's function (the vacuum expectation value of
$\varphi$) is {\it nonzero} \cite{rf33}. Also, the field theory is
renormalizable, and in four dimensions it is asymptotically free and thus
nontrivial \cite{rf34}. Based on these features, we believe that the theory may
provide a useful setting in which to describe the dynamics of the Higgs sector
in the standard model.

Other field theory models whose Hamiltonians are non-Hermitian and
$\mathcal{PT}$-symmetric have also been studied. For example,
$\mathcal{PT}$-symmetric electrodynamics is particularly interesting because it
is asymptotically free (unlike ordinary electrodynamics) and because the
direction of the Casimir force is the negative of that in ordinary
electrodynamics \cite{rf35}. This theory is remarkable because it can determine
its own coupling constant. Supersymmetric $\mathcal{PT}$-symmetric quantum field
theories have also been studied \cite{rf36}.

How does a $g\varphi^3$ theory compare with a $g\varphi^4$ theory? A $g\varphi^3
$ theory has an attractive force. Bound states arising as a consequence of this
force can be found by using the Bethe-Salpeter equation. However, the $g
\varphi^3$ field theory is unacceptable because the spectrum is not bounded
below. If we replace $g$ by $ig$, the spectrum becomes real and positive, but
now the force becomes repulsive and there are no bound states. The same is true
for a two-scalar theory with interaction of the form $ig\varphi^2\chi$, which is
an acceptable model of scalar electrodynamics that has no analog of positronium.
It would be truly remarkable if the repulsive force that arises in a
$\mathcal{PT}$-symmetric quantum field theory having a three-point interaction
could explain the acceleration in the expansion of the universe.

We believe that the concept of $\mathcal{PT}$ symmetry as a generalization of
the usual Dirac Hermiticity requirement in conventional quantum mechanics is
physically reasonable and mathematically elegant. While the proposal of
$\mathcal{PT}$ symmetry is unconventional, we urge the reader to keep in mind
the words of Michael Faraday: ``Nothing is too wonderful to be true, if it be
consistent with the laws of nature.''

\begin{acknowledgments}
I am grateful to the Theoretical Physics Group at Imperial College for its
hospitality and I thank the U.K. Engineering and Physical Sciences Research
Council, the John Simon Guggenheim Foundation, and the U.S.~Department of Energy
for financial support.
\end{acknowledgments}


\begin{enumerate}

\bibitem{rf30} T. T. Wu, Phys. Rev. {\bf 115}, 1390 (1959).

\bibitem{rf2} R. Brower, M. Furman, and M. Moshe, Phys. Lett. B {\bf 76}, 213
(1978); B. Harms, S. Jones, and C.-I Tan, Nucl. Phys. {\bf 171}, 392 (1980) and
Phys. Lett. B {\bf 91}, 291 (1980).

\bibitem{rf1} M.~E.~Fisher, Phys. Rev. Lett.~{\bf 40}, 1610 {1978}; J.~L.~Cardy,
{\it ibid}.~{\bf 54}, 1345 {1985}; J.~L.~Cardy and G. Mussardo, Phys. Lett. B 
{\bf 225}, 275 {1989}; A. B. Zamolodchikov, Nucl. Phys. B {\bf 348}, 619 (1991).

\bibitem{rf7} E. Caliceti, S. Graffi, and M. Maioli, Comm. Math. Phys. {\bf 75},
51 (1980).

\bibitem{rf3} C. M. Bender, K. A. Milton, S. S. Pinsky, and L. M. Simmons, Jr.,
J. Math. Phys. {\bf 30}, 1447 (1989).

\bibitem{rf4} C.~M.~Bender and S.~Boettcher, Phys.~Rev.~Lett. {\bf 80}, 5243
(1998).

\bibitem{rf4.5} Other examples of complex Hamiltonians having $\mathcal{PT}$
symmetry are $H={\hat p}^2+{\hat x}^4(i{\hat x})^\delta$, $H={\hat p}^2+{\hat
x}^6(i{\hat x})^\delta$, and so on (see Ref.~\cite{BBM}). These classes of
Hamiltonians are all {\it different}. For example, the Hamiltonian obtained by
continuing $H$ in (\ref{eq4}) along the path $\delta:\,0\to8$ has a different
spectrum from the Hamiltonian that is obtained by continuing $H={\hat p}^2+{\hat
x}^6(i{\hat x})^\delta$ along the path $\delta:\,0\to4$. This is because the
boundary conditions on the eigenfunctions are different.

\bibitem{rfen} An important technical issue concerns the definition of the
operator $(i{\hat x})^N$ when $N$ is noninteger. This operator is defined in
coordinate space and is used in the Schr\"odinger equation $H\phi=E\phi$, which
reads $-\phi''(x)+(ix)^N\phi(x)=E\phi(x)$. The term $(ix)^N\equiv e^{N\log(ix)
}$ uses the complex logarithm function $\log(ix)$, which is defined with a
branch cut that runs up the imaginary axis in the complex-$x$ plane. This is
explained more fully in Sec.~\ref{sec2}.

\bibitem{rf5} The spectrum of $H={\hat p}^2-i{\hat x}$ is null: I.~Herbst,
Comm.~Math.~Phys.~{\bf 64}, 279 (1979).

\bibitem{foot1} If a system is defined by an equation that possesses a discrete
symmetry, the solution to this equation need not exhibit that symmetry. For
example, the differential equation ${\ddot y} (t)=y(t)$ is symmetric under time
reversal $t\to-t$. The solutions $y(t)=e^t$ and $y(t)=e^{-t}$ do not exhibit
time-reversal symmetry while the solution $y(t)=\cosh(t)$ is time-reversal 
symmetric. The same is true of a system whose Hamiltonian is
$\mathcal{PT}$-symmetric. Even if the Schr\"odinger equation and corresponding
boundary conditions are $\mathcal{PT}$ symmetric, the solution to the
Schr\"odinger equation boundary value problem may not be symmetric under
space-time reflection. When the solution exhibits $\mathcal{PT}$ symmetry, we
say that the $\mathcal{PT}$ symmetry is unbroken. Conversely, if the solution
does not possess $\mathcal{PT}$ symmetry, we say that the $\mathcal{PT}$
symmetry is broken.

\bibitem{rf6} P.~Dorey, C.~Dunning and R.~Tateo, J.~Phys.~A {\bf 34} L391
(2001); {\em ibid}. {\bf 34}, 5679 (2001).

\bibitem{rf8} G.~L\'evai and M.~Znojil, J.~Phys. A{\bf 33}, 7165 (2000).

\bibitem{rf9} B.~Bagchi and C.~Quesne, Phys.~Lett. A{\bf 300}, 18 (2002).

\bibitem{rf10} Z.~Ahmed, Phys.~Lett. A{\bf 294}, 287 (2002); G.~S.~Japaridze,
J.~Phys.~A{\bf 35}, 1709 (2002); A.~Mostafazadeh, J.~Math.~Phys. {\bf 43}, 205
(2002); {\em ibid}. {\bf 43}, 2814 (2002); D.~T.~Trinh, PhD Thesis, University
of Nice-Sophia Antipolis (2002), and references therein.

\bibitem{rf11} C.~M.~Bender and A.~Turbiner, Phys. Lett. A {\bf 173}, 442
(1993).

\bibitem{rf12.5} C. M. Bender, S. Boettcher, H. F. Jones, and V. M. Savage,
J.~Phys.~A: Math.~Gen.~{\bf 32}, 6771 (1999).

\bibitem{rf12} C.~M.~Bender and S.~A.~Orszag, {\it Advanced Mathematical Methods
for Scientists and Engineers}, (McGraw-Hill, New York, 1978), Chaps.~3 and 10.

\bibitem{BBM} C.~M.~Bender, S.~Boettcher, and P.~N.~Meisinger,
J.~Math.~Phys. {\bf 40}, 2201 (1999).

\bibitem{rf14} C.~M.~Bender, D.~C.~Brody, and H.~F.~Jones, Phys.~Rev.~Lett.
{\bf 89}, 270401 (2002) and Am.~J.~Phys. {\bf 71}, 1095 (2003).

\bibitem{rf13} A. Mostafazadeh, J.~Math.~Phys.~{\bf 43}, 3944 (2002).

\bibitem{rf15} C.~M.~Bender, S.~Boettcher, and V.~M.~Savage, J.~Math.~Phys. {\bf
41}, 6381 (2000); C.~M.~Bender, S.~Boettcher, P.~N.~Meisinger, and Q.~Wang,
Phys.~Lett. A {\bf 302}, 286 (2002).

\bibitem{rf16} G. A. Mezincescu, J.~Phys.~A: Math.~Gen.~{\bf 33}, 4911 (2000);
C.~M.~Bender and Q.~Wang, J.~Phys. A: Math. Gen. {\bf 34}, 3325 (2001).

\bibitem{rf17} P.~A.~M.~Dirac, Proc.~R.~Soc.~London A {\bf 180}, 1 (1942).

\bibitem{rf18} C.~M.~Bender, P.~N.~Meisinger, and Q.~Wang, J.~Phys.~A: Math.
Gen. {\bf 36}, 1973 (2003).

\bibitem{foot2} If a function satisfies a linear ordinary differential equation,
then the function is analytic wherever the coefficient functions of the
differential equation are analytic. The Schr\"odinger equation (\ref{eq9}) is
linear and its coefficients are analytic except for a branch cut at the origin;
this branch cut can be taken to run up the imaginary axis. We choose the
integration contour for the inner product (\ref{eq17}) so that it does not cross
the positive imaginary axis. Path independence occurs because the integrand of
the inner product (\ref{eq17}) is a product of analytic functions.

\bibitem{rf20} C.~M.~Bender, M.~V.~Berry and A.~Mandilara, J.~Phys.~A: Math.
Gen. {\bf 35}, L467 (2002).

\bibitem{rx} C.~M.~Bender, D.~C.~Brody, and H.~F.~Jones, Phys.~Rev.~D
{\bf 70}, 025001 (2004).

\bibitem{rf20.5} C.~M.~Bender and H.~F.~Jones, Phys.~Lett.~A {\bf 328}, 102
(2004).

\bibitem{rf23} C.~M.~Bender and G.~V.~Dunne Phys. Rev. D {\bf 40}, 2739 and 3504
(1989).

\bibitem{DY} F. J. Dyson, Phys. Rev.~{\bf 85}, 631 (1952).

\bibitem{rf31} T. Hollowood, Nucl. Phys. B {\bf 384}, 523 (1992).

\bibitem{rf32} C.~M.~Bender, G.~V.~Dunne, and P.~N.~Meisinger,
Phys. Lett. A {\bf 252}, 272 (1999).

\bibitem{rf33} C.~M.~Bender, P.~Meisinger, and H.~Yang,
Phys. Rev. D {\bf 63}, 45001 (2001).

\bibitem{rf34} C.~M.~Bender, K.~A.~Milton, and V.~M.~Savage,
Phys.~Rev.~D~{\bf 62}, 85001 (2000).

\bibitem{rf35} C. M. Bender and K. A. Milton,
J.~Phys.~A: Math. Gen. {\bf 32}, L87 (1999).

\bibitem{rf36} C.~M.~Bender and K.~A.~Milton,
Phys. Rev. D {\bf 57}, 3595 (1998).

\end{enumerate}
\end{document}